\documentclass[preprint]{aastex631}

\newcommand\Romn{{\color{orange}{\it Roman}}}
\newcommand\Romns{{\color{orange}{\it Roman's}}}
\newcommand\Rubn{{\color{cyan}{\it Rubin}}}
\newcommand\Rubns{{\color{cyan}{\it Rubin's}}}
\newcommand{\okina}{\textquoteleft}
\newcommand{\Ou}{{\okina}Oumuamua}

\makeatletter
\def\namedlabel#1#2{\begingroup
    #2%
    \def\@currentlabel{#2}%
    \phantomsection\label{#1}\endgroup
}
\makeatother

\graphicspath{{./}{figures/}}

\begin{document}

\title{R2--D2: Roman and Rubin -- from Data to Discovery}

\clearpage

\collaboration{9}{The AURA Roman--Rubin Synergy Working Group}
\author{Suvi Gezari ({\it chair, STScI/JHU})} 
\author{Misty Bentz ({\it Georgia State})}
\author{Kishalay De ({\it MIT})}
\author{K. Decker French ({\it UIUC})}
\author{Aaron Meisner ({\it NOIRLab})}
\author{Michelle Ntampaka ({\it STScI/JHU})}
\author{Robert Jedicke ({\it Hawai`i})}
\author{Ekta Patel ({\it UC Berkeley})}
\author{Daniel Perley ({\it Liverpool})}
\author{Robyn Sanderson ({\it UPenn})}
\collaboration{21}{Outside Experts}
\author{Christian Aganze ({\it UC San Diego})}
\author{Igor Andreoni ({\it Maryland/JSI})}
\author{Eric F. Bell ({\it Michigan})}
\author{Edo Berger ({\it Harvard})}
\author{Ian Dell'Antonio ({\it Brown})}
\author{Ryan Foley ({\it UC Santa Cruz})}
\author{Henry Hsieh ({\it Planetary Science Institute})}
\author{Mansi Kasliwal ({\it Caltech})}
\author{Joel Kastner ({\it Rochester Institute of Technology})}
\author{Charles D. Kilpatrick ({\it Northwestern/CIERA})}
\author{J. Davy Kirkpatrick ({\it IPAC})}
\author{Casey Lam ({\it UC Berkeley})}
\author{Karen Meech ({\it Hawai`i})}
\author{Dante Minniti ({\it Universidad Andr\'{e}s  Bello})}
\author{Ethan O. Nadler ({\it Carnegie/USC})}
\author{Daisuke Nagai ({\it Yale})}
\author{Justin Pierel ({\it STScI})}
\author{Irene Shivaei ({\it Arizona})}
\author{Rachel Street ({\it LCOGT})}
\author{Erik J. Tollerud ({\it STScI})}
\author{Benjamin Williams({\it University of Washington})}

\section{Executive Summary}
The NASA Nancy Grace Roman Space Telescope (\Romn) and the Vera C. Rubin Observatory Legacy Survey of Space and Time (\Rubn), will transform our view of the wide-field sky, with similar sensitivities, but complementary in wavelength, spatial resolution, and time domain coverage.  
 Here we present findings from the AURA \Romn$+$\Rubn\ Synergy Working group\footnote{https://outerspace.stsci.edu/display/RRS/AURA+Roman+Rubin+Synergies+Working+Group}, charged by the STScI and NOIRLab Directors to identify frontier science questions in General Astrophysics, beyond the well-covered areas of Dark Energy and Cosmology, that can be uniquely addressed with \Romn\ and \Rubn\ synergies in observing strategy, data products and archiving, joint analysis, and community engagement.  This analysis was conducted with input from the community in the form of brief (1-2 paragraph) ``science pitches'' (see Appendix), and testimony from ``outside experts'' (included as co-authors).

\Rubn\ will provide wide, deep, multi-band seeing-limited optical imaging with a regular cadence of observations.  \Romn\ will provide wide, deep, multi-band diffraction-limited near-infrared imaging with grism and prism low-resolution spectroscopic capabilities.  The core science goals, and technical requirements to achieve these goals, have been defined separately for \Romn\ \footnote{https://www.stsci.edu/roman/documentation/technical-documentation}
and \Rubn\ \footnote{https://www.lsst.org/scientists/keynumbers}.  However, substantial effort has been made towards examining how these surveys could be combined to improve our measurements of Dark Energy and Dark Matter, e.g., improvements in photometric redshifts, weak-lensing shear measurements, the galaxy power spectrum, galaxy cluster selection, time-delays in strong-lensing systems, and type Ia supernova distances \citep{Jain2015, chary_jsp}.  Here we expand further into areas of General Astrophysics, and identify a rich and broad landscape of potential discoveries catalyzed by the combination of exceptional quality and quantity of \Romn\ and \Rubn\ data, and summarize implementation requirements that would facilitate this bounty of additional science with coordination of survey fields, joint coverage of the Galactic plane, bulge, and ecliptic, expansion of General Investigator and Target of Opportunity observing modes, co-location of \Romn\ and \Rubn\ data, and timely distribution of data, transient alerts, catalogs, value-added joint analysis products, and simulations to the broad astronomical community.

\section{Milky Way and Solar System Science}
\normalsize


\subsection{What is the composition, physical state, and morphology\\of the small bodies in the solar system?}

The Sun, Moon, comets and five planets have been observed and studied by humans since the time they became a distinct species, but new populations of objects in our solar system continue to be discovered.  Within the past few decades our understanding of our solar system's formation and evolution has been revolutionized by e.g. the discovery of the trans-Neptunian objects, the irregular satellites of giant planets and, just within the past few years, the detection of the first two macroscopic interstellar objects (ISOs) passing through the inner solar system.  These new classes of objects probe the edges of our understanding of the formation of our and exoplanetary systems, while the extreme members in color, physical state, and morphology of long-known populations test the limits of our ability to model the solar system's evolution.


One of \Rubns\ key science goals is to generate a catalog of the solar system that will contain an order of magnitude more objects than are currently known. Within its vast set of discoveries \Rubn\ is likely to identify extreme objects within and passing through our solar system such as main belt object cratering or disruption events, small objects with outrageous morphologies, near-Earth objects on impact trajectories, interstellar objects, and perhaps even a distant planet in our own solar system. \Romn\ will provide detailed observations of these faint, extreme objects to explore how they fit into the puzzle of our own solar system's formation and the larger picture of all planetary systems. High resolution \Romn\ images of `active' small objects will determine whether the events are due to thermal spin 
up, impacts, or other processes. High precision \Romn\ photometry will allow the determination of the shapes of the faintest and most elongated objects in the solar system e.g. to determine whether 1I/Oumuamua's unusual shape remains uncommon when compared to similar-size objects in our solar system.  

\begin{figure}[!ht]
\centering
\includegraphics[width=0.49\textwidth]{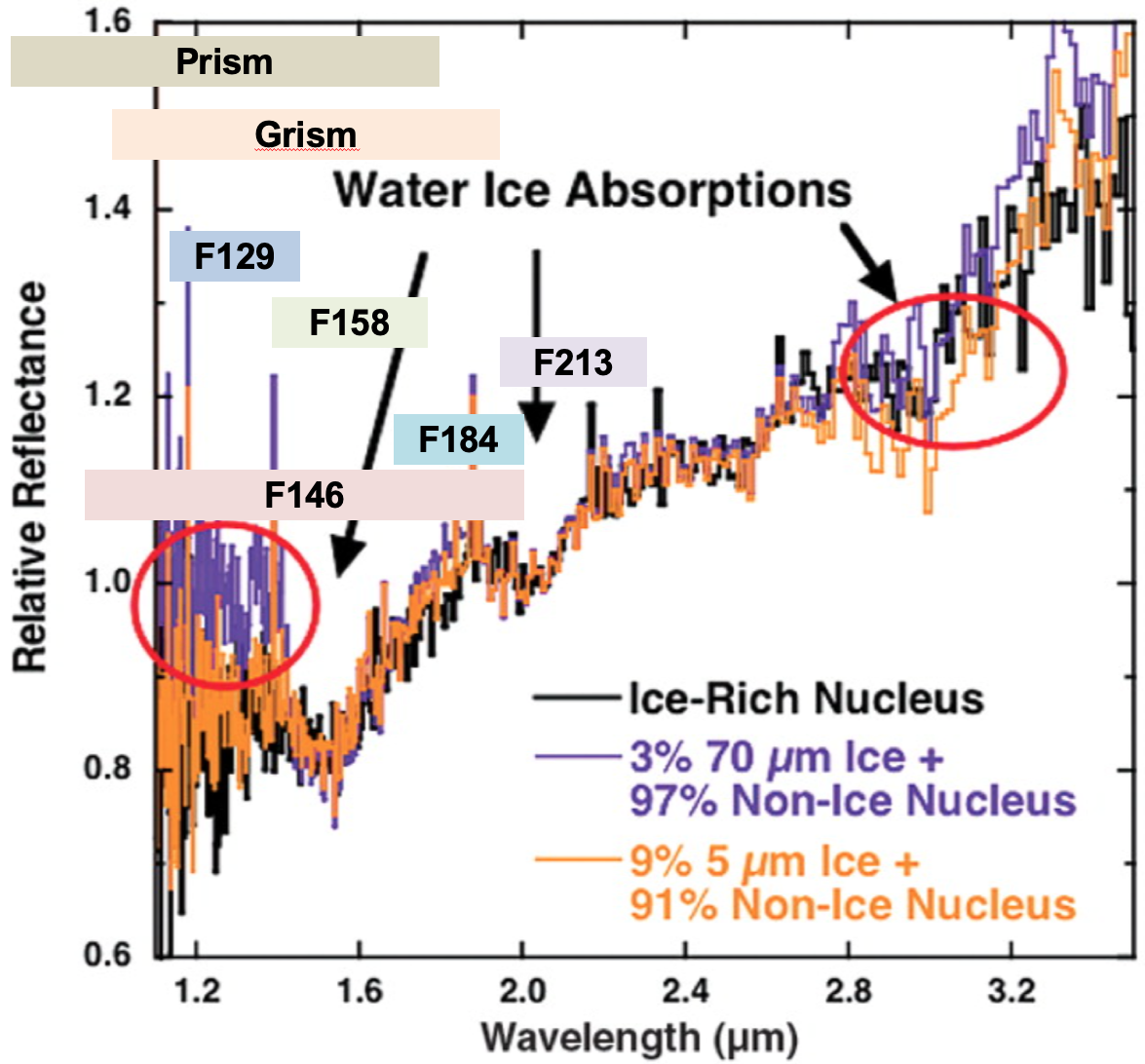}
\caption{Spectrum showing evidence for solid water ice deposits on the surface of comet 9P/Tempel 1,  including a representation of \Romns\ prism, grism, and filter pass bands.  \Romns\ imaging filters, and spectroscopic coverage and resolution, are well suited to small body targets and identifying water ice absorption features at about $1.6\mu$m and $2.0\mu$m.  Figure from \citealt{Sunshine-2006-WaterIceOnTempel}, modified to include \Romn\ passbands.}
\label{fig:CometaryWaterIceFeatures}
\end{figure}

\Romn\ and \Rubn\ together will provide spectrophotometric characterization of interesting targets from the near-UV through the NIR while \Romns\ prism and grism will e.g. enable identifcation of H$_2$O absorption features in asteroids and comets.  If an object is detected on an Earth-impact trajectory, \Romn's faint limiting magnitude and space-based vantage point will provide the necessary astrometry to refine its orbit and impact probability. Finally, the LSST provides the best opportunity in the foreseeable future of detecting a distant planetary-size object in our own solar system \citep{Sheppard2016, Batygin2016}. \Romn's unique capabilities will be important to revealing its properties and establishing its dynamical and evolutionary trajectory to its current state and location.

\Rubns\ all-sky surveying, repeat cadence, and deep limiting magnitude are an ideal combination for detecting the extreme members of the population of small objects within and passing through our solar system.  For instance, \Rubn\ will detect asteroid collisions in the main belt \citep[e.g.][]{Bottke-2005-Main-Belt-Collisions}; interstellar objects passing through our solar system \citep[e.g.][]{Meech-2017-Oumuamua}; objects with extremely slow and fast rotation rates \citep[e.g.][]{Pravec2000-AsteroidRotation};  those with remarkable colors \citep[e.g.][]{Sheppard2010-ExtremeTNOs}; distant objects on cometary orbits that have not yet begun to sublimate volatiles \citep[e.g.][]{Jewitt-2005-Damocloids} and, perhaps, unknown planets bound to our Sun \citep[e.g.][]{Mustill201-ExoplanetInSolarSystem}.  

\citealt{Holler-2018-RomanSolarSystemScience} and \citealt{Holler-2021-RomanSolarSystemScience} provide an exhaustive discussion of the small body science potential with \Romn\ and synergistic opportunities with \Rubn.  The two systems will operate in different but overlapping wavelength regimes that will enable calibrated spectrophotometry from the near-UV to the near-IR to determine the composition of the most compelling small bodies identified in our solar system.  \Romns\ NIR capability makes it a powerful tool for characterizing active and disrupted objects because it will be sensitive to larger ejected dust particles that persist longer in a comet's and asteroid's environs than smaller dust particles.

Some of the objects detected by \Rubn\ will be too faint for followup characterization by ground-based telescopes and their transient nature will require rapid acquisition and characterization with \Romns\ imaging filters, grism and prism.  For instance, the first ISO, 1I/\Ou, was discovered on 2017 Oct 19 and the last ground-based observations occurred on Nov 23, but observations with {\it Hubble} doubled the time range of observations and were critical to identifying its non-gravitational acceleration \citep{Micheli-2018-Oumuamua-nongrav}.

One of the more compelling solar system science problems is measuring the distribution of water using  small bodies as tracers.  Some asteroids show evidence of hydration in an absorption feature at about $0.9\mu$m \citep[e.g.][]{Rivkin-2002-HydratedMineralsOnAsteroids}.  \Romns\ low-resolution prism instrument is well-suited to identifying and characterizing this broad band while the prism, grism and some of the filters will be important for identifying water-ice absorption features near $1.6\mu$m and $2.0\mu$m (Figure \ref{fig:CometaryWaterIceFeatures}).

See Table of Implementation Requirements: \ref{non-sidereal},    \ref{ecliptic}, \ref{roman_mo},  \ref{mpc},  \ref{cross-match-moving}, \ref{precovery}.



\subsection{What are the demographics and formation histories of stars and substellar objects in the Milky Way?}

One of \Romn\ Space Telescope's core scientific goals is to improve our understanding of exoplanet demographics through a dedicated microlensing campaign (\Romn\ Galactic Bulge Time Domain Survey). \Romn\ and \Rubn\ have the potential to teach us much more about the prevalence and formation of planetary and substellar mass objects, beyond the baseline \Romn\ program. How does accretion work during the star/planet formation processes? How does the low-mass initial mass function (IMF) vary as a function of environmental parameters such as metallicity? How prevalent are free-floating substellar objects compared to those orbiting higher-mass stars? These questions can be addressed through suitable combination of \Romn\ and \Rubn\ observations/capabilities in the following areas: (1) detection and characterization of substellar objects via photometry and microlensing, and (2) exploiting eruptive YSOs to understand the temporal evolution of accretion during star and planet formation.

\begin{figure}[!hbt]
    \centering
     \setlength{\belowcaptionskip}{-5pt}
    \includegraphics[width=0.45\textwidth]{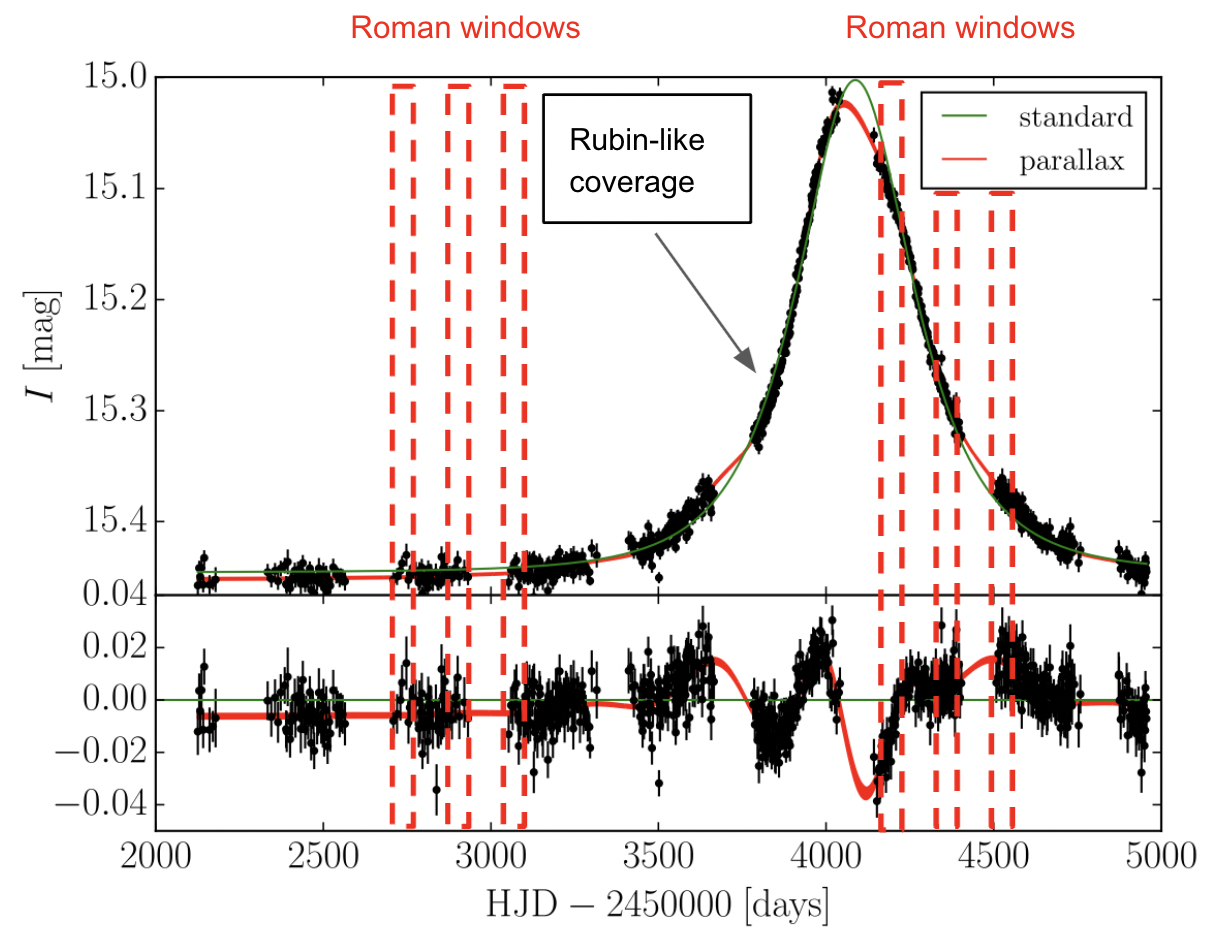}
    \caption{Candidate $8.7M_{\odot}$ BH microlensing event OGLE3-ULENS-PAR-02 from the OGLE survey. The black dots show the observable cadence with a ground-based survey like \Rubn\ to derive a microlensing parallax model (red line), while the red dashed boxes show expected \Romn\ observing windows to observe the astrometric microlensing shift.  Figure from \citet{Wyrzykowski2016}.}
    \label{fig:compact_objects}
\end{figure}

At substellar masses, for example free-floating brown dwarfs, our knowledge of the initial mass function is limited by the small volume over which existing surveys can detect/discover these cold objects \citep{kirkpatrick_20pc}. The combination of \Romn\ and \Rubn\ data at red-optical and near-infrared wavelengths will dramatically expand the volume over which we can detect L and T type brown dwarfs \citep[e.g.,][]{stauffer_astro2020, aganze21}. This will allow us for the first time to begin probing the low-mass initial mass function in the Galactic thick disk and halo, and thereby check for any metallicity dependence of the low-mass star formation process \citep[e.g.,][]{bromm2013, chabrier2014} [Meisner, Sainio pitches]. As shown in Figure \ref{fig:compact_objects}, \Romn\ monitoring of the \Romn\ Galactic Bulge Time Domain Survey footprint will suffer from temporal gaps in coverage. By filling in these gaps, \Rubn\ can help constrain models of \Romn\ Galactic Bulge Time Domain Survey microlensing events, pushing to relatively high masses. Simultaneous \Rubn\ monitoring during \Romn\ Galactic Bulge Time Domain Survey observations can also break exoplanet microlensing degeneracies. [Street pitch]

\begin{figure}[!htb]
    \centering
    \includegraphics[width=0.43\textwidth]{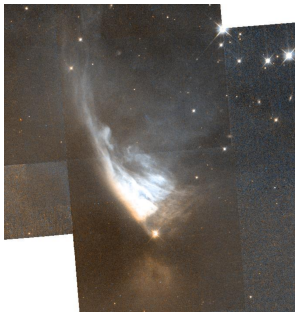}
    \caption{High resolution Hubble Space Telescope ({\it Hubble}) imaging from the Hubble Legacy Archive of the nearby YSO PV Cep in outburst, which is one of the handful of objects with such imaging. The image reveals evidence of a conical nebula carved out by an outflow. \Romn\ high resolution imaging follow-up of $\Rubn$-discovered YSO outbursts will disambiguate the sources of the eruptions and establish the presence and morphologies of cavities carved by outflows from these YSOs.}
    \label{fig:yso}
\end{figure}

In terms of further understanding star and planet formation, \Rubn\ (through its wide area, sensitivity and repeat coverage) will transform the study of young stellar object (YSO) outbursts \citep[e.g.,][]{audard2014,elbakyan2021} by detecting them in greater numbers and to larger distances (in more varied environments) than ever before. Such outbursts may account for a significant fraction of accreted mass during star formation. Pre-existing \Romn\ imaging (to characterize the quiescent state) and follow-up (both imaging and spectroscopy) will be needed to understand the physical properties of erupting YSOs (Figure \ref{fig:yso}) identified from \Rubn. Although YSO outburst decay time scales can be up to years or even decades, obtaining prompt follow-up shortly after the eruption is highly valuable for constraining models of these events. [Ginsburg, Kastner pitches]

See Table of Implementation Requirements:  \ref{too_roman}, \ref{too_roman_spec},  \ref{rges_rubin}, \ref{plane_roman},  \ref{cross-match},    \ref{precovery}.

\vspace{10pt}




\subsection{How do compact objects form and evolve in the Milky Way?}

As end products of the most massive stars through mass loss and powerful supernova (SN) explosions, the demographics of neutron stars (NSs) and black holes (BHs) are central to our understanding of stellar evolution. While NSs have a hard surface, the gravitational influence of BHs is {\it the only way to find and characterize them}. Historically, BHs have been discovered primarily via X-ray emission from surrounding hot material accreted from a binary companion (e.g. \citealt{Blackcat}). We are now entering an era where multiplexed spectrographs like DESI and SDSS-V, and astrometric missions like {\it Gaia} can search for the long-sought population of non-accreting BHs in wide binaries via radial velocity (e.g. \citealt{Gu2019, Wiktorowicz2020}) and astrometric (e.g. \citealt{Mashian2017, Yamaguchi2018}) shifts, respectively. Yet, not only is our understanding of the accreting NS/BH population demographics severely incomplete (e.g. \citealt{Gandhi2020, Jonker2021}), we have just beginning to detect some of the estimated $\sim 10^7-10^8$ isolated Galactic BHs \citep{Fender2013, Olejak2020}, or the 25,000 BHs expected to have clustered in the central parsec of the Galactic Center \citep{Miralda2000}. 

\subsection{Roman and Rubin synergies into the dark universe: Isolated compact objects}

Microlensing is the only way to characterize free-floating Galactic BHs \citep{Paczynski1986}. With $\sim 10^4$ microlensing events expected during its Bulge survey \citep{Penny2019}, \Romn\ could find $\sim 100$ BHs and $\sim 300$ NSs \citep{Gould2000a, Lam2020} that would offer a completely new window into the remnant mass function. As intrinsically massive objects, NSs/BHs are expected to be characterized by long-lived events \citep{Wiktorowicz2019} spanning much longer than the planned $\approx 72$ day observing seasons. A \Rubn\ survey of the \Romn\ bulge footprint to fill-in the gaps between the observing seasons (Figure \ref{fig:compact_objects}) would be crucial for photometric coverage to constrain the microlensing parallax \citep{Gould2000b}, in addition to enhancing the overall rate of planets, substellar and stellar systems discovered via microlensing [Street pitch].  Also, unlike ground based surveys, \Romn\ would be uniquely placed to detect the astrometric microlensing signal \citep{wfirst2019} and offer exquisite constraints on the lens mass to systematically search for these NS/BH lenses [Lu, Lam, Wyrzykowski pitches], as well as determine the masses for free-floating planet candidates. This has now been demonstrated with {\it Hubble} follow-up observations of candidate BH microlensing events from the ground-based OGLE survey, with the confirmation of astrometric microlensing by a 7.1 M$_\odot$ BH \citep{Sahu2022} as well as a catalog of several dark compact objects \citep{Lam2022}.

See Table of Implementation Requirements:  \ref{rges_rubin},  \ref{rges_rubin_pre}.


\begin{figure*}[!htb]
\centering
\includegraphics[width=0.80\textwidth]{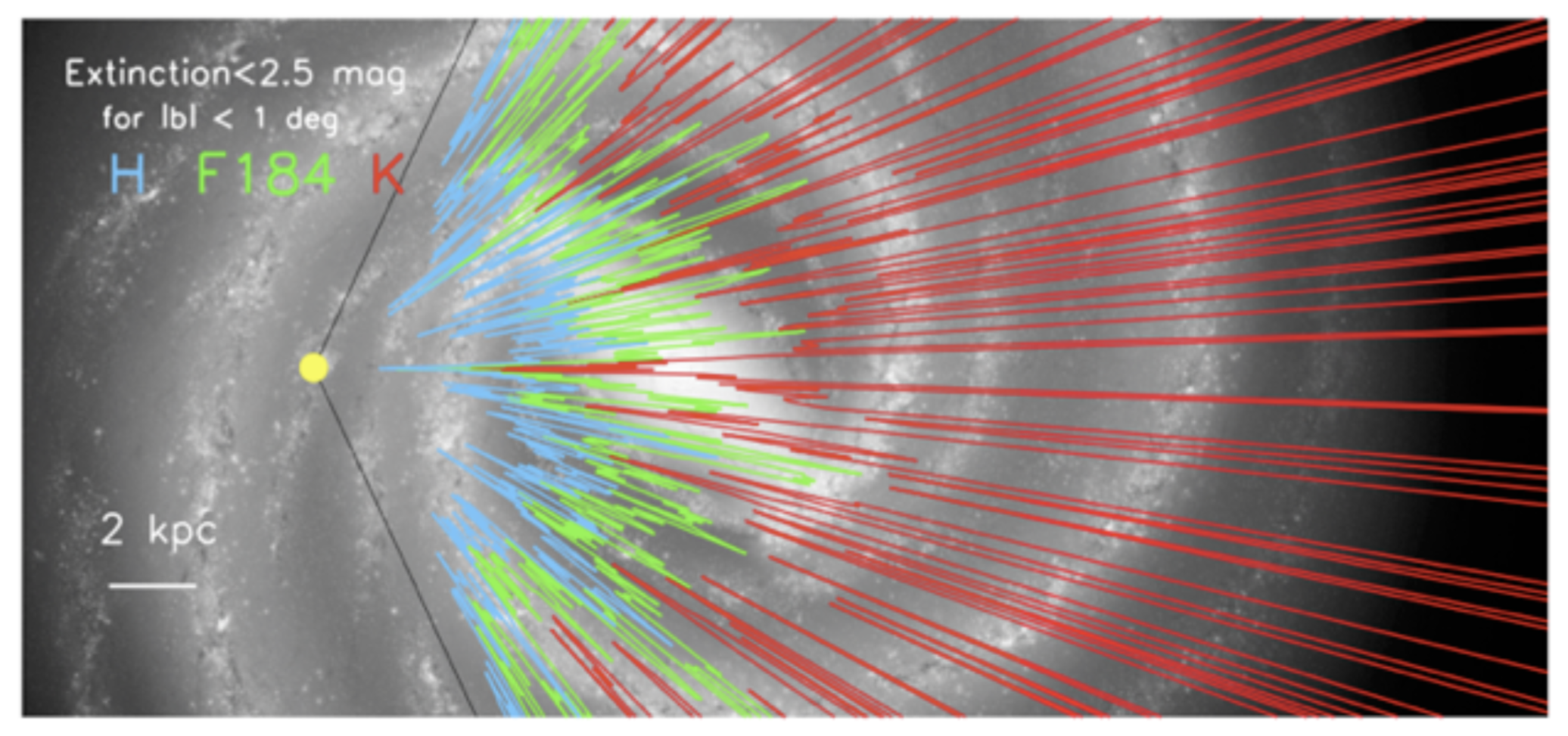}
\caption{The extinction ``horizon" in the H, F184 and K bands in the Galactic plane. \Romn\ and \Rubn\ can together map stars and dust throughout essentially the entire Milky Way plane, providing an unprecedented three-dimensional view of our Galaxy (figure from \citealt{helou_k_band}; visual by Robert Hurt).}
\label{fig:helou_kband}
\end{figure*}

\subsection{Roman and Rubin synergies into the dark universe: Accreting compact objects}

With its southern location, the \Rubn\ observatory will be capable of creating an exquisite time domain movie of outbursting sources in the Galactic plane — dwarf novae, novae and X-ray binaries \citep{Strader2018}. \Romns\ planned high cadence time domain survey of the Galactic bulge will provide an excellent complement to \Rubns\ capabilities by providing unprecedented historical photometry of the progenitors, allowing us to map out the poorly understood roles of these eruptive episodes on their long term evolution \citep{Johnson2019, Maccarone2019}. \Romns\ Galactic bulge survey will also detect the most dust obscured events and completely new types of transient phenomena, where \Rubn\ limits will be very constraining. A wide area \Romn\ map of the \Rubn\ Galactic plane footprint could routinely enable the characterization (e.g. distance, evolutionary state)  of the progenitor systems, which remain poorly studied due to their very faint quiescent counterparts.

See Table of Implementation Requirements:  \ref{plane_roman},  \ref{cross-match}.



\subsection{What is the distribution of stars and dust in our Galaxy?}

What is the distribution of stars and dust throughout the Milky Way, and what does this tell us about our Galaxy's formation and evolution? Does the Milky Way have well-defined spiral arms, and what would it look like to an external observer? SDSS enabled detailed Milky Way tomography of (mostly) extragalactic fields \citep[e.g.,][]{juric_sdss_tomography}, revealing for the first time large-scale substructures in the Galactic distribution of stars, for instance the Sagittarius and Monoceros streams \citep{Newberg2002}. Pan-STARRS, covering 3/4 of the sky and most of the Galactic plane, took this one step further, allowing for novel three-dimensional maps of stars and dust \citep[e.g.,][]{green_ps1}. {\it Gaia} has now complemented Pan-STARRS with high-accuracy trigonometric distances over the entire sky, another major step toward making the ultimate three-dimensional map of stars and dust in our Milky Way Galaxy \citep[e.g.,][]{green_with_gaia, helmi2018, antoja2018}.



However, {\it Gaia} and Pan-STARRS are fundamentally limited in terms of their depths and wavelength coverage, which does not extend beyond $\sim$1~$\mu$m. This lack of depth and infrared sensitivity limits the extent to which the Galactic plane's distant and/or heavily dust-obscured reaches can be mapped.

The combination of very deep \Rubn\ optical and \Romn\ near-infrared photometry throughout the Galactic plane can make an unprecedented map of the 3D distribution of stars and dust in dense regions of the Milky Way. \Romns\ angular resolution is highly complementary to \Rubn\ in terms of the improved deblending of \Rubn\ sources that it can offer. \Romn\ near-infrared photometry of the plane can push the 3D map to distances/reddenings at which \Rubn\ has lost sensitivity (see Figure \ref{fig:helou_kband}). [Zucker pitch]

Focusing more specifically on the Milky Way bulge/nucleus, the combination of \Romn\ and \Rubn\ can complete the RR Lyrae  census in this central region \citep[e.g.,][]{minniti_bulge_rrl}, thereby unraveling mysteries about how the nuclear star cluster was assembled: did it form through merging of primordial globular clusters? How old is the Galactic nuclear star cluster? Multi-epoch \Romn/\Rubn\ data can also reveal clusters, streams and kinematics of old stellar populations near the Galactic center. For all of these Milky Way mapping applications, \Romns\ combination of large field of view plus {\it Hubble}-like angular resolution and depth uniquely enables this science. [Minniti, Fern\'{a}ndez-Trincado pitches]

See Table of Implementation Requirements:  \ref{bulge_rubin},  \ref{plane_roman},  \ref{cross-match},  \ref{forced}.
 




\section{Transient Discovery and Transient Hosts}

\normalsize

\subsection{Using gravitationally lensed SNe as a time machine and an independent measure of $H_0$.}

One of the most intriguing developments to arise in the era of ``precision cosmology" is the tension between the value of Hubble's constant of expansion ``$H_0$" measured from direct local measurements, and the model-dependent value inferred from the Cosmological Microwave Background. Understanding if this tension is real, and not due to systematic errors or model assumptions, is critical for providing evidence for new physics beyond the standard model.  Gravitationally lensed SNe provide a direct and completely independent measure of $H_0$ from these two methods.  

\begin{figure}
    \centering
    \includegraphics[scale=0.2]{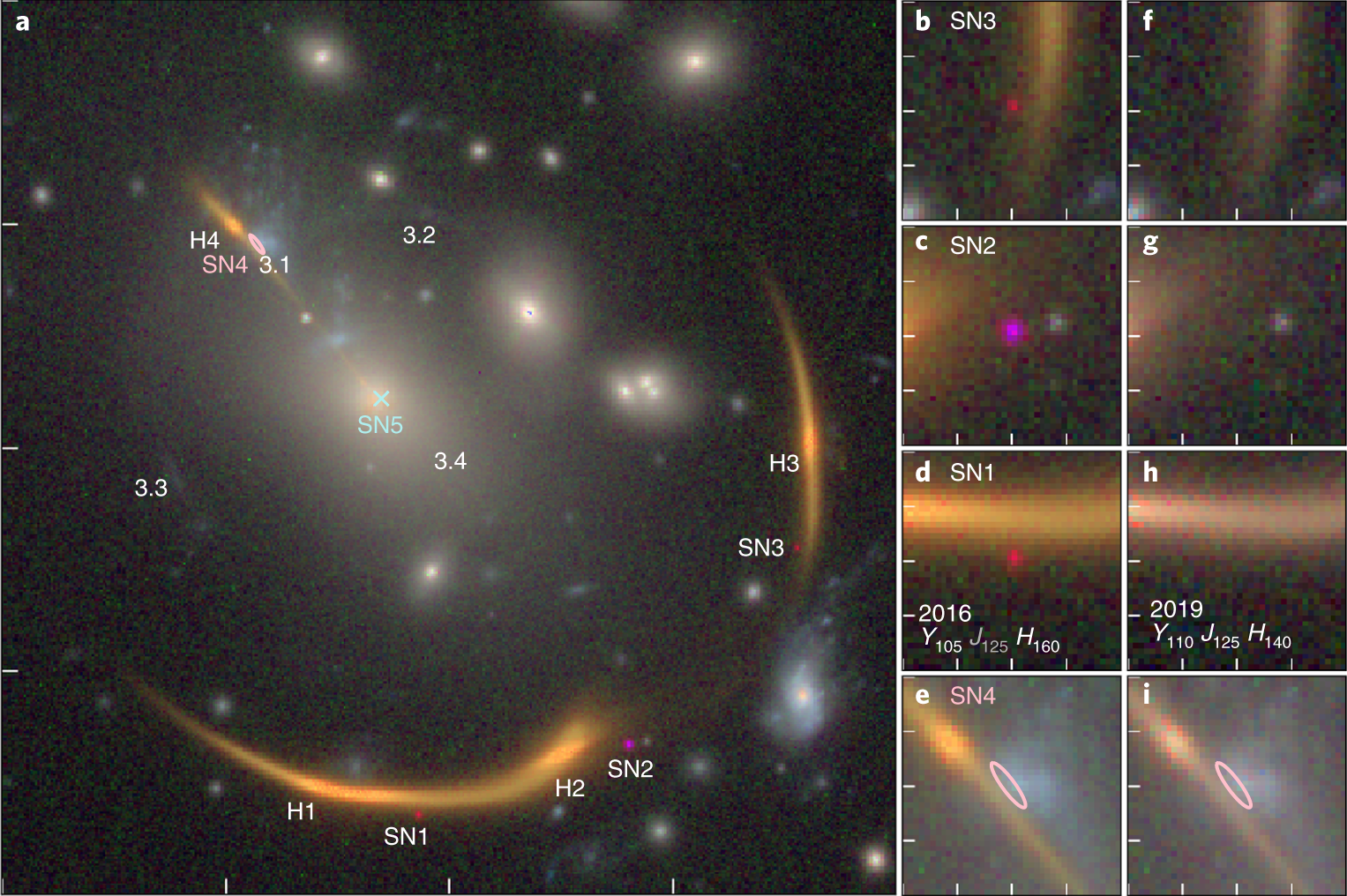}
    \caption{{\it Hubble} images of the lensed SN AT 2016jka ($z$=1.95) in the galaxy cluster lens system MACS J0138.0-2155 ($z$=0.338).  Follow-up space-based imaging from \Romn\ of lensed SN discovered by \Rubn\ is required in order to resolve the multiple images of the lensed SN, and measure their relative time delays.  Figure from \citealt{Rodney2021}.}
    \label{fig:my_label}
\end{figure}

A SN that explodes in a background galaxy that is strongly lensed by a foreground galaxy or cluster, resulting in multiple images due to gravitational bending of the SN light, can be used to measure $H_0$ by measuring the relative time delays between each SN image.  SNe have several advantages for so-called ``time delay cosmography" compared to other variable phenomena, such as variable quasars, including the fact that their intrinsic light curves and luminosities are well determined, especially for the class of Type Ia SNe.  This method has been attempted in the only three known multiply imaged SNe: Type II SN Refsdal \citep{Kelly2015}, Type Ia SN 2016geu \citep{Goobar2017}; Type Ia SN AT2016jka \citep{Rodney2021}.  While each of these objects had unique challenges that limited their cosmological utility, high quality observations for dozens of such systems could, in principle, yield constraints on $H_0$ to better than 1\% precision \citep{Birrer2021}.  Furthermore, one can use the magnification and time delays of the lensed SN images as a ``time machine", to study a SN in its earliest explosive phase \citep{Shu2018}.  This will be particularly valuable for SNe in cluster lens systems where the time delays are on the order of $\sim 100$ days, giving time to predict the date and location of a subsequent SN image to shed light on the SN progenitor and pre-explosion mass-loss, as well as probing deviations from a smooth lensing model for the foreground mass distribution \citep{Goobar2017}.

\begin{figure}
    \centering
    \includegraphics[scale=0.55]{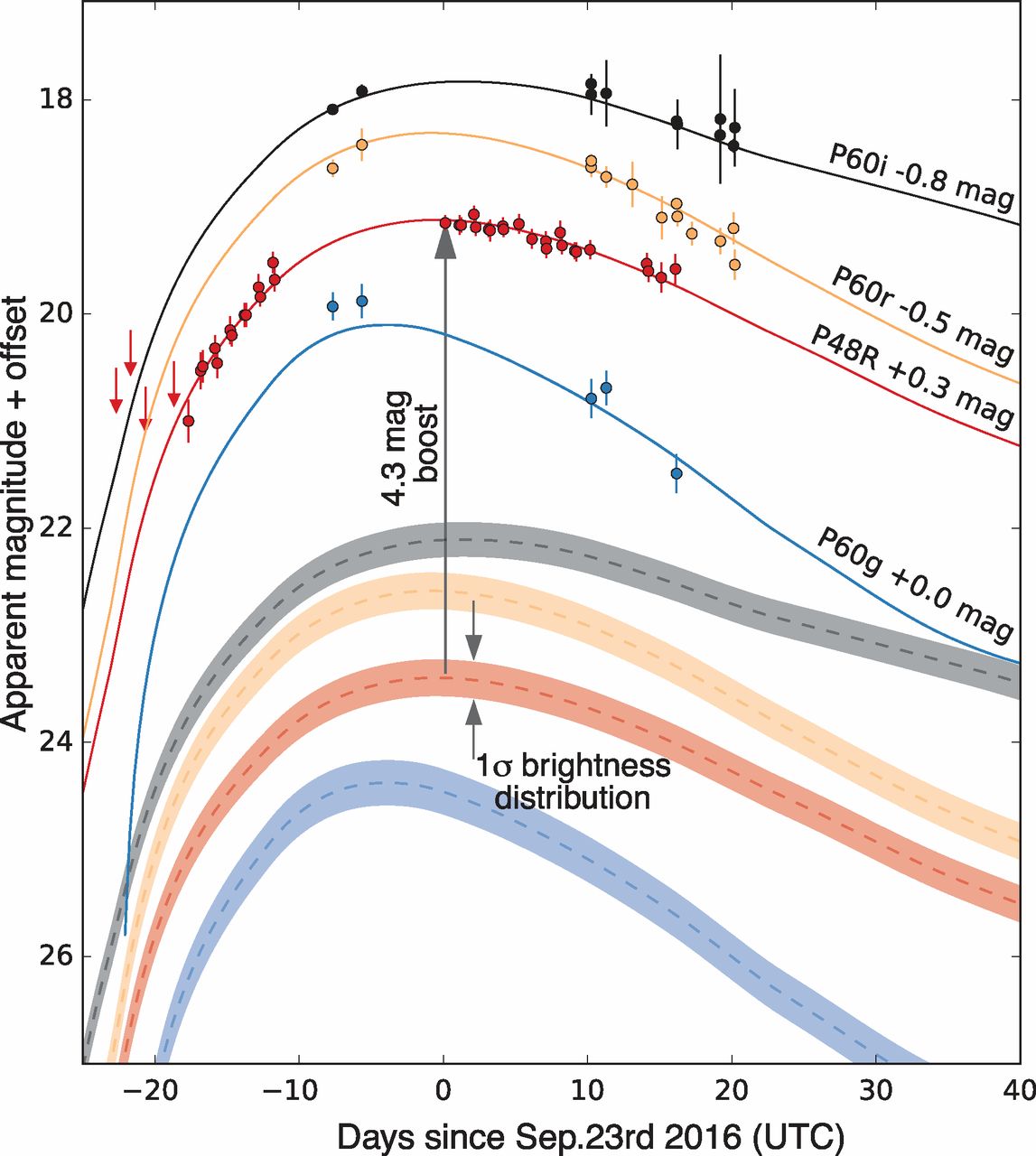}
    \caption{Figure from \citealt{Goobar2017} which demonstrates how a lensed type Ia SN was identified from being overluminous in its ground-based light curve measured by the Palomar Transient Factory (PTF).  A similar method could be applied to spectroscopically confirmed SNe (or photometrically identified SNe with a host-galaxy redshift or photo-z) discovered by \Rubn.}
    \label{fig:snIalens}
\end{figure}

A large enough sample of lensed SNe, could also be used as a competitive probe of dark energy.  Fortunately, the rates of strongly lensed SNe expected from \Rubn\ (hundreds: \citealt{Oguri2010}) and \Romn\ (tens: \citealt{Pierel2021}) as calculated from mock catalogs, are reaching the levels relevant for such experiments.  \Romn\ and \Rubns\ lens samples will also be highly complementary in redshift, with \Romn\ best suited to detect higher-redshift lensed SNe ($z > 0.5$) \citep{Pierel2021}.  There are several possible observational methods for finding and monitoring lensed SNe, from blind time domain searches by looking for an overluminous SNe (Figure \ref{fig:snIalens}), to monitoring of known galaxy-galaxy or galaxy-cluster strong lens systems \citep{Shu2018}.  
\Romns\ high angular resolution imaging can be used to identify arcs surrounding the lensing galaxy, and multi-color imaging is essential to maximize the contrast of the lens arc and model the magnification. [Holwerda pitch]

Identification of strong lenses at scale can be done with machine learning, a citizen science approach or from spectroscopic signatures of the blend of galaxies in a galaxy-galaxy lens system. A vetted list of strong lenses can then be monitored as a ``watch list" in the \Rubn\ transient alert stream for the signature of a SN going off in the source galaxy.  With a pre-calculated catalog of photo-z's, when a candidate SN is detected in \Rubn\ with an overluminous peak absolute magnitude, spectroscopic follow-up can confirm the SN redshift and trigger follow-up imaging with \Romn\ to confirm the multiple images.
 
See Table of Implementation Requirements:  \ref{rubin_roman_tda}, \ref{ml_roman},  \ref{watchlist_rubin}.



\subsection{How do galaxies and black holes co-evolve through cosmic time?}

\begin{figure}
    \centering
    \includegraphics[scale=0.25]{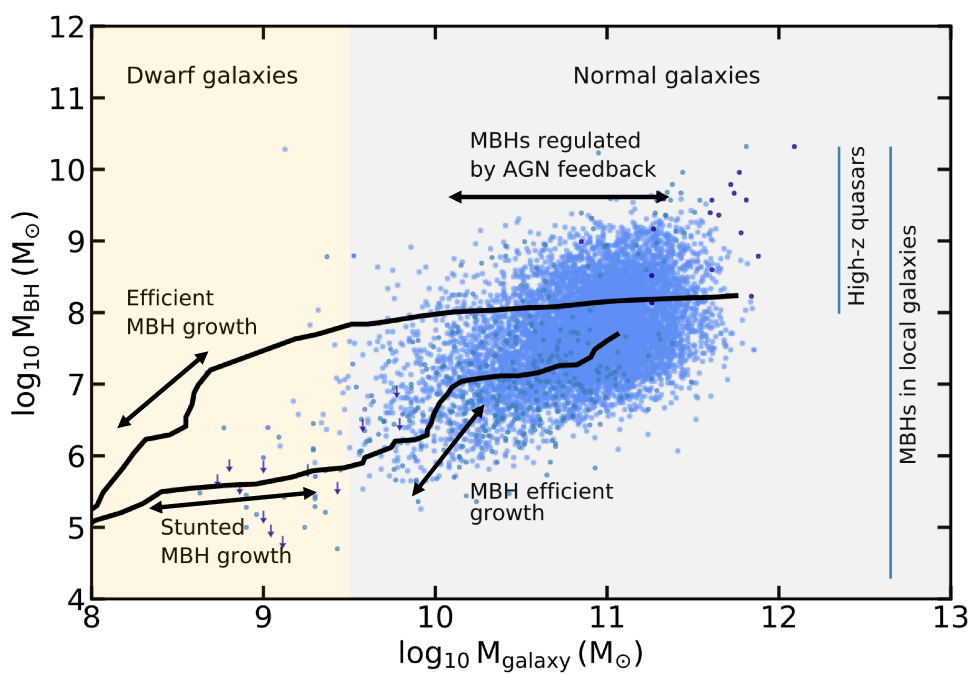}
    \caption{Comparison between observations and simulations showing the growth of massive black holes within their host galaxies. From \cite{volonteri21}.}
    \label{bhgal_evol}
\end{figure}

The co-evolution of massive black holes and the galaxies that host them is supported by several observational and computational lines of evidence, yet the details of the physical mechanisms driving this symbiotic relationship are still uncertain.
A fundamental question in the context of galaxy and black hole co-evolution is the nature of the black hole seeds (see the recent review by \citealt{volonteri21}).  Observational evidence for $\sim10^9$ solar mass black holes when the universe is $<1$\,Gyr old \citep{fan01} creates difficulties for models of galaxy evolution in which remnants from the first stars coalesce to form the central black hole seed.   More massive black hole seeds from, for example, direct collapse of density fluctuations in the early universe can grow to these large sizes in a shorter time, yet we have no observational evidence to support such massive seeds nor the existence of the conditions required for them to form and grow. [Bentz pitch] The wide field, time domain nature of \Rubn\ will allow for the identification of low- or intermediate-mass black holes in the local universe via the detection of stochastic nuclear variability in dwarf galaxies \citep{Baldessare2018, Ward2021}, as well as previously dormant black holes in quiescent galaxies that reveal themselves by tidally disrupting and accreting a star \citep{Gezari2021}.  The high resolution imaging, and wide area coverage of \Romn\ will characterize the host galaxies of these accreting central black holes, including luminous quasars [Lacy pitch].  Thus \Rubn\ and \Romn\ together can map the properties of black holes and their host galaxies over a large range of mass scales, providing important constraints on the formation mechanisms of seeds in the early universe, and their growth through accretion and mergers over cosmic time (see Figure \ref{bhgal_evol}).

Black hole feeding, while easy to identify, still presents numerous challenges to our understanding of how gas is funneled from large galactic scales down into the nucleus and ultimately onto the black hole.  The smallest relevant spatial scales cannot be resolved with imaging, and so time domain techniques such as reverberation mapping (or echo mapping) have been developed \citep{cackett21}.  Using time resolution rather than spatial resolution to study light echoes in AGN allows their internal structures to be mapped.  Yet the number of AGN that have been studied in this way remains small because of the typical necessity of targeting individual objects and monitoring them over long time scales.  Reverberation mapping studies of a large number of AGN spanning a more diverse range of black hole and galaxy properties is needed to improve our understanding of black hole feeding [Homayouni, Lyu, Shen pitches].  \Rubn\ light curves may provide an important component to facilitate such studies.  If a time domain component is also provided by \Romn\, then the mapping of AGN structures on spatially-unresolvable scales becomes possible through the combination of \Romn\ and \Rubn\ photometry (dust torus reverberation) and/or \Romn\ grism spectroscopy plus \Rubn\ photometry (emission line reverberation).

Finally, mergers provide a separate but important pathway for the growth of massive black holes.  Galaxy mergers are relatively straightforward to identify and much progress has been made in understanding how the process unfolds, yet to date there are few observational constraints on the merger process between their massive black holes \citep{deRosa19}.  High-resolution images of galaxies hosting dual AGN with separations ranging from scales of kpc to sub-arcsec are needed to inform the pairing and evolution of the merger process for massive black holes. [Shen, Liu pitches]  \Rubn\ monitoring will uncover dual or blended AGN through their variability, while high resolution \Romn\ imaging will constrain their apparent separations as well as the merger stage of the host.

See Table of Implementation Requirements:   \ref{GI_roman}, \ref{GI_roman_spec}, \ref{south},   \ref{rubin_roman_tda},   \ref{cross-match}.



\subsection{How do the rates and properties of all classes of transients depend on their host galaxy environments and stellar populations?}

Observations of transient events provide brief windows to study stars, compact objects, and their underlying environments. Combining observations from \Romn\ and \Rubn\ will enable studies of transient environments on scales ranging from the integrated host galaxy, to localization within the host galaxy, to identification of the progenitor star. These studies will deepen our understanding of known transients and help to interpret the range of new transients that will be discovered with \Romn\ and \Rubn.

By combining photometry from \Romn\ and \Rubn, properties of transient host galaxies can be determined to help interpret known transients and to more efficiently prepare for detecting new transients [French pitch]. The stellar masses, star formation rates, recent star formation histories, photometric redshifts, structural parameters, color gradients, and levels of dust extinction can be measured using the combined \Rubn\ $ugrizY$ and \Romn\ $0.5-2\mu$m photometry for transient host galaxies. Spectroscopic classification will only be possible for a small sub-set of the transients discovered with \Rubn\ or \Romn. The host galaxy properties of a transient can be used to interpret the likely class of the transient and to identify the most interesting transients for follow-up observations. Several studies have found that given the difference in progenitor systems for type Ia SNe and core-collapse SNe, the differing stellar populations in the host galaxies allow for transients to be classified based on the host galaxy information alone \citep{Baldeschi2020,Gagliano2021}. Tidal disruption events (TDEs) occur at high rates in post-starburst, green valley, and centrally-concentrated galaxies \citep{Arcavi2014,French2016,LawSmith2017,Graur2018,Hammerstein2021,Arcavi2021}. Similarly, the host galaxies of changing-look AGN \citep{Dodd2021} have also been found to lie preferentially in green valley host galaxies, and may provide clues on what is driving their extreme changes in accretion rate [Bentz pitch]. Using photometry from \Romn\ and \Rubn\, especially making use of \Romns\ spatial resolution to identify highly concentrated galaxies, likely host galaxies of interesting nuclear transients can be pre-identified, such that new transients can be rapidly flagged for multi-wavelength followup. 

On more local scales, resolving the location of a transient within a galaxy can provide important information about the nature of a transient [Gezari pitch]. Even a single \Romn\ observation of a transient would provide an improved localization constraint compared to the information available from \Rubn. Localization with \Rubn\ will be limited to $\sim$0.1 arcsec at best, whereas \Romn\ can localize transients to few$\times10^{-3}$ arcsec. If a transient can be constrained to be consistent with a galaxy nucleus on milliarcsecond scales, contamination from SNe can be all but removed when searching for TDEs or other SMBH transients. Similarly, the localization of a transient to a spiral arm or star forming region within a galaxy indicates the transient is likely to be associated with the death of a massive star. Good localization constraints can enable more detailed comparisons of SNe with the metallicities and stellar populations of their parent environment \citep[e.g.][]{Galbany2018}. Type Ic SNe are found in more metal-rich environments and super-luminous SNe (SLSNe) are found in more metal-poor environments \citep{Galbany2018,Izzo2018}. These trends are obscured by the varying conditions across a galaxy, especially when comparing the central bulge -- which may dominate the light in a fiber spectrum -- and the conditions in the outer disk.

\begin{figure}
    \centering
    \includegraphics[scale=0.25]{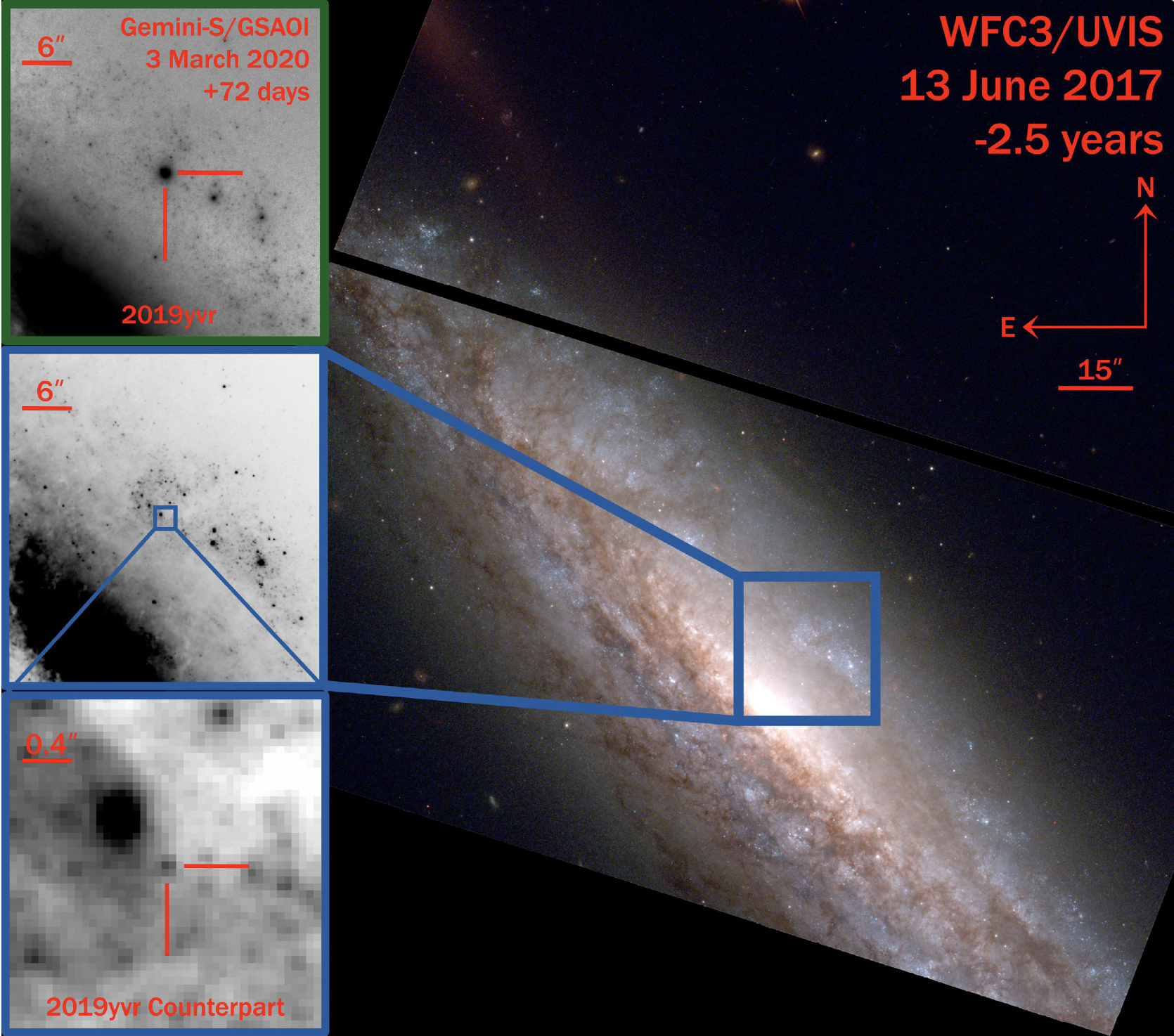}
    \caption{Example of the massive progenitor star of the Type Ib SN 2019yvr, discovered in pre-explosion {\it Hubble} imaging.  Figure from \citet{Kilpatrick2021}. }
    \label{fig:sn_progenitor}
\end{figure}

On the smallest scales, progenitor stars of SNe imaged years or decades before the explosion can provide direct links from massive star progenitors to the types of SNe we observe \citep{Podsiadlowski1993,Smartt2009}. We show a recent example of the likely progenitor for SN 2019yvr in the host galaxy NGC 4666 from \citet{Kilpatrick2021} in Figure \ref{fig:sn_progenitor}. While we cannot predict the locations of all future SNe, many SNe are discovered in nearby galaxy clusters such as Virgo and Fornax every year, even in the era before \Rubn\ and \Romn. A targeted \Romn\ survey of these nearby clusters would provide deep pre-explosion imaging of SN to be discovered in the years after the targeted survey [Kilpatrick pitch]. The IR imaging capabilities of \Romn\ are well-suited to investigate the progenitors of SNe in dusty star forming regions or cases of failed SN where the progenitors are obscured in the optical by circumstellar dust.

See Table of Implementation Requirements:  \ref{south},  \ref{roman_wide},  \ref{rubin_roman_tda}, \ref{roman_cluster}, \ref{cross-match}, \ref{sed}, \ref{cross_galaxy}.



\subsection{What are the electromagnetic properties of gravitational wave events, their relation to binary properties, and implications for r-process nucleosynthesis?}
\label{sec:gw}

To address this question we require rapid, deep, and dedicated optical/NIR follow-up of gravitational wave events and short-duration gamma-ray bursts (SGRBs).  In the case of GW-detected mergers, the sky localizations are expected to be at best $\sim 10$ deg$^2$ and the distances will be tens to hundreds of Mpc.  The resulting faintness of the optical/NIR emission may therefore necessitate the use of \Rubn\ and/or \Romn, and would benefit from mutual coordinated follow up to assess the presence of both optical and NIR emission.  For example, in the case of the only existing GW-EM detected merger to date (GW170817) there appear to be two emission components: one peaking in the optical on a timescale of $\sim{\rm day}$ and the other peaking in the NIR on a timescale of $\sim{\rm week}$.  Whether this is ubiquitous to NS-NS mergers (or present in NS-BH mergers) is a key question that will help to address the origin of these two components.  

In the case of SGRBs, the mergers are viewed along the axis of the relativistic jet and the emission is therefore significantly brighter at early time. Such events are predominantly located at higher redshifts than GW-detected events, up to $z\sim 1$, and they therefore provide a complementary sample of NS-NS/NS-BH mergers to what the GW detectors can find.  This sample can enable studies of redshift evolution in the properties of the mergers, as well as a different angular view (face-on orientation in SGRBs compared to all viewing angles in GW-detected mergers). However, the majority of SGRBs are localized poorly (detections by {\it Fermi}), with similar localizations to GW events. \Romn$+$\Rubn\ rapid coordinated follow-up of these events will increase the SGRB detection rate by at least a factor of three, and will more than double the NS-NS merger rate when combined with GW-detected events.  

Another promising avenue for the detection of NS-NS and NS-BH merger counterparts is the "serendipitous" discovery of kilonovae in wide-field survey data, i.e. independently of external GW or SGRB triggers. These sources will likely be found at large distances beyond the horizon of the ground-based GW detectors. They will allow us to study the emission at all viewing angles, correlations with redshift, and possibly enable yet another type of cosmological studies \citep{Coughlin2020}. Recent works have shown that a few tens of kilonovae are expected to be discovered by the Rubin main survey \citep{Setzer2019, Cowp2019, Andreoni2022}. Remarkably, Roman will have the capability to detect kilonovae out to redshift z~1 \citep{Chase2021}.

The key scientific questions that will uniquely be addressed by \Romn$+$\Rubn\ synergy are: (i) How prevalent is $r$-process nucleosynthesis in NS-NS/NS-BH mergers? What are the typical ejecta masses and velocities?  How do these quantities depend on the properties of the merging objects (NS vs.~BH, masses, mass ratios)? (ii) What is the angular structure of the ejected matter, and what can this teach us about the merger process? (iii) Do all mergers produce relativistic jets? (iv) Is there significant evolution in the nature of the binaries and the properties of the ejected matter as a function of redshift (at $z\sim 0-1$)?


The key requirement for implementation is rapid target-of-opportunity (TOO) follow up with both \Romn\ and \Rubn\, covering the larger localization regions to a sufficient depth to detect (or rule out) an optical/NIR counterpart (kilonova or afterglow). In practice, this means observations to as faint as $\sim 24$ AB mag (and potentially fainter for more distant sources) over an area of up to tens of deg$^2$.  We generally expect the optical emission to peak first, so rapid response with \Rubn\ may need to lead the rapid response with \Romn.  However, since we have only one example of an optical/NIR counterpart of a GW-detected merger, we do not yet know the relation between optical and NIR emission for the population.  Note that in \citet{Andreoni2021} they show that with proper planning, TOO observations with \Rubn\ would only interrupt $\sim 1\%$ of the baseline survey.  The capabilities of TOO observations with \Romn\ have also been investigated. However, theoretically we expect the IR to be more ubiquitous for kilonovae than the optical - it is sensitive to a much wider range in viewing angle, mass ratio, lanthanide fraction and remnant lifetime. In fact, some NS-NS and NS-BH mergers may not even be discoverable by \Rubn\ and only be accessible to \Romn, making TOO NIR imaging and spectroscopy follow-up with \Romn\ even more critical (see Kasliwal et al.~whitepaper in Appendix of \citealt{Kasliwal2013}).

See Table of Implementation Requirements:  \ref{too_roman},  \ref{too_rubin},  \ref{comm_mma}.

\subsection{What end-states of stars exist in nature that have yet to be discovered?}
\label{sec:endstates}

The mapping between the lives of stars (as we observe them in our Milky Way Galaxy and nearby galaxies) and the deaths of stars (as we observe them as SNe or other transients, and as remnants) remains only crudely sketched.  The textbook picture is that a star's death is controlled primarily by its initial mass, but this simple picture has slowly been unravelling in recent years.  Observations have been a driving force in driving our understanding of stellar death, in particular the appreciation that the outcomes of stellar evolution are much more diverse than what appears in textbooks.  The discovery of SNe that are extremely luminous or extremely feeble, those that are unusually long-lived or extremely short-lived, those with unusual spectral properties and so on have all provided new questions about the evolution and end-states of stars.  This revolution remains incomplete due to the limited capabilities of current facilities, which have limited depth and largely only scan the optical range.

\begin{figure}
    \centering
    \includegraphics[scale=0.45]{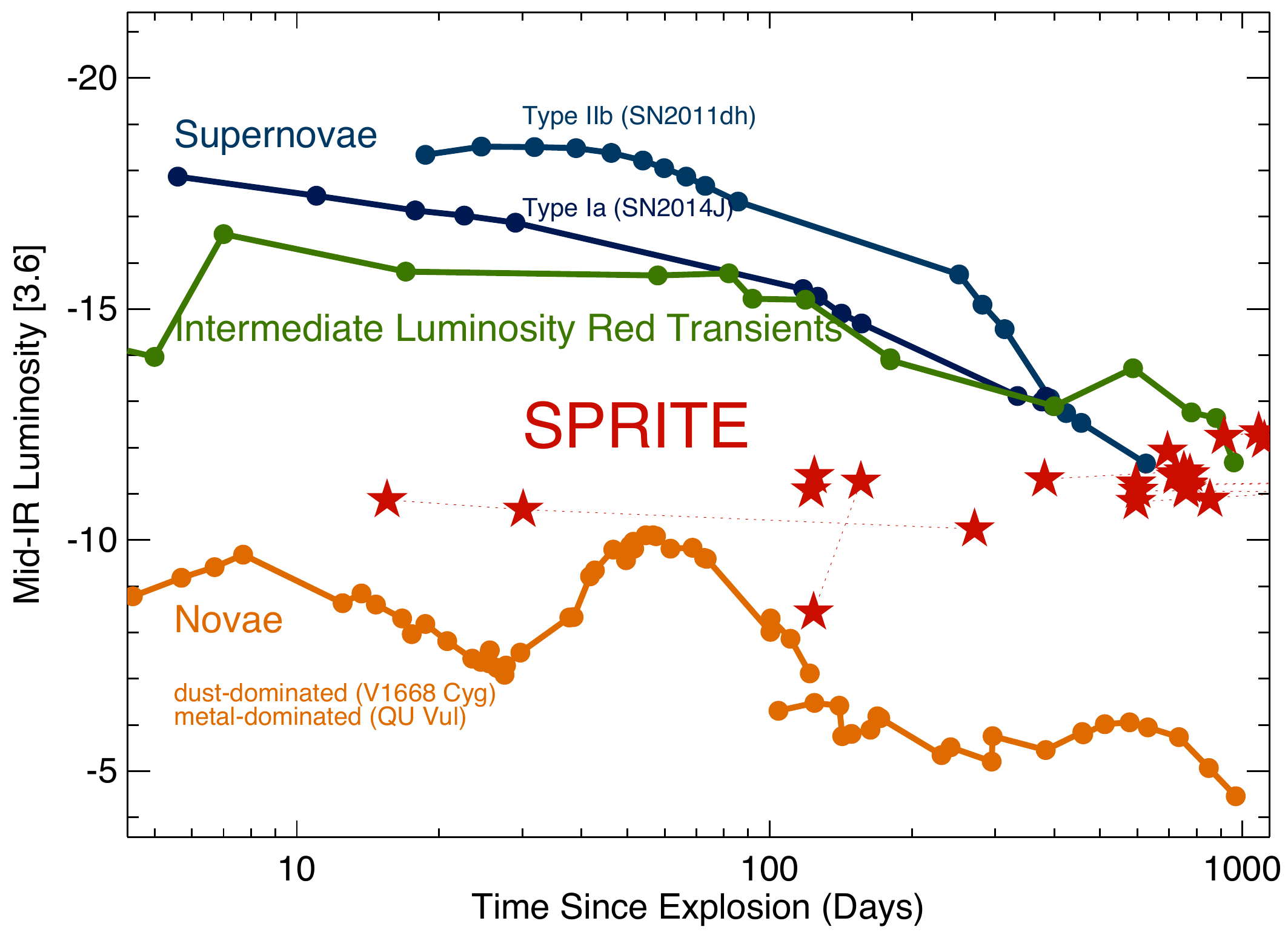}
    \caption{Spitzer Infrared Intensive Transients Survey discovery of a new class of IR transients known as SPRITEs (eSPecially Red Intermediate-luminosity Transient Events), in comparison to previously known classes of IR-bright transients.  Figure from \citet{Kasliwal2017}}
    \label{fig:sprites}
\end{figure}

\Romn\ and \Rubn\ together provide the ability to cover major ``blind spots'' of current surveys. \Romns\ infrared survey capabilities allow it to survey for transients that are concealed by dust---whether that dust originates from the surrounding galaxy environment or from the stellar progenitor.  This gives it a unique ability to probe a range of phenomena not accessible to optical surveys.  In the central regions of LIRGs and ULIRGs, star-formation is at its most furious and the potential for unusual forms of stellar death due to enhancement of the high-mass IMF or dynamical interactions are maximized.  It also has the power to survey for self-obscured transients, for which the optical display is shielded by shells of circumstellar dust synthesized by the progenitor.  Any other end-of-life stellar outcome with a characteristic temperature that causes it to emit primarily in the infrared range will also be uniquely detected by \Romn---kilonovae (case \ref{sec:gw}) are the classic example, but there may be other phenomena that peak in this region, as suggested by a recent Spitzer survey (see Figure \ref{fig:sprites}), including electron-capture SNe from very low-mass stars, and mergers of massive stars \citet{Kasliwal2017}.   Finally very high-redshift transients from early generations of stars will inevitably be redshifted in the the infrared range.

The synergy with \Rubn\ is critical to this effort, because the phenomena above would otherwise be lost in a sea of ordinary optical SNe without the benefit of deep optical limits and without accurate photometric redshift constraints.  \Rubn\ observations close in time with \Romn\ ones will allow genuinely red transients to be picked out and the \Rubn\ deep reference imaging will permit rapid estimates of their luminosity and distance, which are key to appropriately targeting follow-up with other facilities.


See Table of Implementation Requirements:  \ref{south},  \ref{roman_wide}, \ref{rubin_roman_tda},   \ref{alerts_roman}.



\section{Galaxies, Large-Scale Structure, and Dark Matter}
\normalsize

\subsection{What is the 3D distribution of small scale structure?}

The dozens of ultra-faint dwarf (UFD) galaxies and stellar streams located at distances up to $\sim$400 kpc from the Milky Way and Andromeda (M31) have dramatically augmented our understanding of galaxy formation and evolution at small scales. The continuous stream of precise radial velocity and 6D phase space information for MW and M31 halo stars are yet another rich dataset with which we can outline the evolutionary history of the Local Group’s two most massive galaxies. However, the extent of our knowledge of hierarchical galaxy formation on the smallest scales is generally rooted in observations of only these two galaxy halos. Understanding the MW and M31 in a cosmological context requires a statistically significant sample of similar mass analogs and a census of their halo substructures at depths and wide-field areas that only \Romn\ and \Rubn\ can reach. 

\begin{figure}
    \centering
    \includegraphics[scale=0.45]{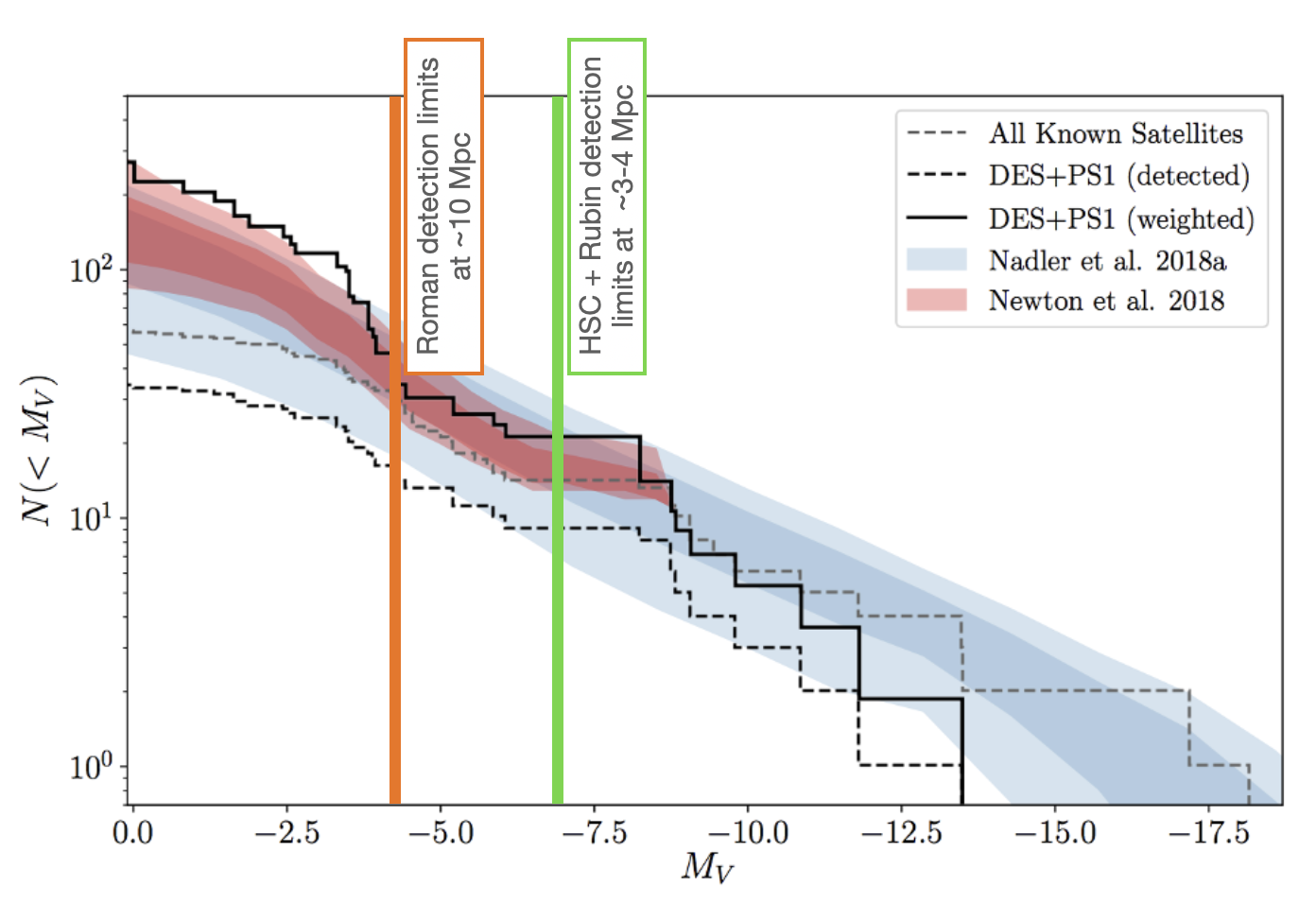}
    \caption{Observed and predicted galaxy luminosity function for the Milky Way adapted from \citet{Drlica-Wagner2020}. The green vertical line shows approximate estimates for discovery potential with HSC+\Rubn\ at $\sim$3-4 Mpc. The orange line indicates predictions for \Romn\ at $\sim$10 Mpc.}
    \label{fig:galaxies}
\end{figure}

The \Rubn\ observatory is roughly expected to discover all MW satellites with $M_V < 0$ mag, $\mu < 32$ mag/arcsec$^2$ in the Southern hemisphere, and it will nearly double the number of known satellite galaxies around nearby galaxies ($\sim$3-4 Mpc) to put them on roughly equal footing with our current knowledge of M31’s galaxy luminosity function \citep[$M_V \sim -6.5$; E. Bell, Fig. \ref{fig:galaxies}; e.g.,][]{drlicawagner19, mutlupakdil21}. It will also yield the most thorough search for stellar streams in the outer halo ($>$ 50 kpc) of the MW [Belokurov pitch]  and around nearby galaxies \citep{Pearson2019}. \Rubn\ will also enable the detection and characterization of stellar halos around dwarf and Milky-Way mass galaxies allowing for direct comparison to numerical simulations of hierarchical assembly [Williams pitch]. Together, dwarf galaxies, streams and stellar halos provide three unique probes to (i) reconstruct the dynamical history of the MW and analogous nearby galaxies, (ii) constrain the properties of dark matter halos ($\sim 10^6 - 10^{12} \, M_{\odot}$), and (iii) to test models of galaxy formation at small scales \citep[see][]{BullockBK17}. 

Owing to its dramatically improved star-galaxy separation and deep point-source limits, \Romn\ will be more sensitive still to dwarf galaxies, streams and stellar halos. \Romn\ will enable follow-up searches of satellite galaxies around nearby hosts ($\sim$3-4 Mpc), probing nearly 10x deeper ($M_V\sim-4$ mag), yielding 10s$-$100s of dwarfs per galaxy system \citep{bell_wfirst}. With a statistical sample of host galaxies and their corresponding satellite populations at these faint limits, \Romn$+$\Rubn\ will allow us to draw the first statistical connections between the faintest known satellite populations and their hosts, including (i) correlations between the presence of a massive satellite and satellite phase space configurations \citep[e.g.,][]{Battaglia22,GaravitoCamargo21, Patel20} and (ii) the subsequent imprints that satellite galaxies leave throughout their hosts' stellar halos, including stellar wakes and streams \citep[e.g.,][]{GaravitoCamargo19}. 

Such a large sample of UFDs across a variety of environments will also provide the most robust constraints on the formation, evolution, and properties of UFDs (i.e., initial mass functions), the oldest known stellar systems in the Universe \citep{Simon2019}. Furthermore, UFDs and any potential evidence for perturbations and gaps in streams will illuminate the threshold for galaxy formation in subhalos, star formation in the early Universe, and further our understanding of reionization (i.e., the near/far connection, \citealt{Weisz2019}). \Romn$+$\Rubn\ will also reveal thousands of strong lens measurements \citep{Oguri2010, Collett2015, Weiner2020} that, along with precise measurements of UFD population statistics, will sharpen constraints on dark matter properties that influence the abundance and properties of small subhalos, including its primordial temperature, interactions, and particle mass \citep{nadler21a,nadler21b} [Nadler, Bechtol pitches]. 

\begin{figure}
    \centering
    \includegraphics[scale=0.7]{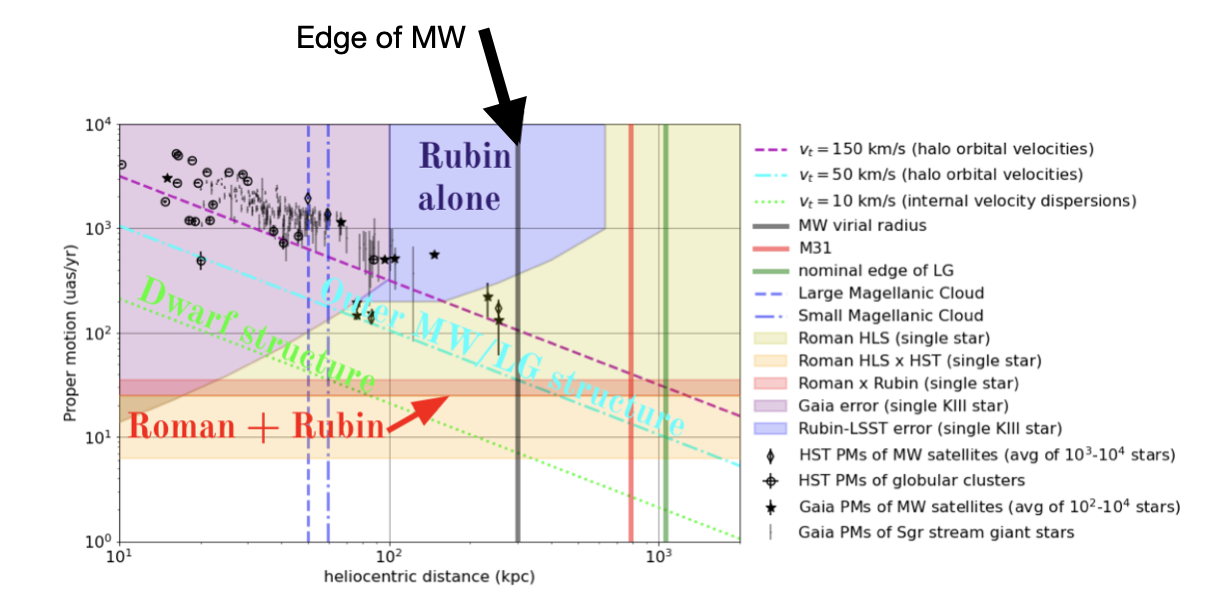}
    \caption{Proper motion accuracies for different astrometric surveys, relative to positions and proper motions of various local-group objects. Figure adapted from \citet{Sanderson2019}.}
    \label{fig:astrom}
\end{figure}

Fitting a simple Gaussian-process model of turbulence to reconcile the sky positions of point sources measured by a space-based telescope with their locations in ground-based surveys can drastically improve astrometry from the ground. Using {\it Gaia} positions and Dark Energy Survey images in the riz bands, \citet{fortino21} reduced astrometric variance due to atmospheric turbulence by a factor of about 12, resulting in RMS turbulence errors of $\sim$3-5 mas per exposure. Assuming 20 exposures over a nominal 5-year baseline this results in an expected proper-motion precision of $\sim$35 uas/yr with current data. This translates to a tangential velocity uncertainty of $\sim$50 km/s at $\sim$300 kpc from the Galactic center, comparable to measured orbital speeds of typical MW objects at that distance. Improvement by a factor $\sim$5 would allow us to determine precise orbits to the virial radius, while an improvement by a factor $\sim$10 would allow us to measure internal velocity dispersions of dwarfs (Figure \ref{fig:astrom}).

The biggest improvement in accuracy for this type of joint reduction comes from an increased density of astrometric sources \citep{fortino21}. \Romn\ will provide much more densely sampled point-source data since a typical \Romn\ exposure is both 4-6 mag deeper than {\it Gaia} and far less affected by extinction. \Romn\ will be an exquisite astrometric instrument since the core-science bulge microlensing survey will deliver detailed knowledge of the point-spread function that is necessary for the success of the core-science weak lensing survey at high latitudes. In addition, \Romn\ will benefit from using {\it Gaia’s} full extended-mission catalog as its astrometric frame (WFIRST AWG 2018). Thus we can expect \Romn\ to have far higher source density at similar astrometric accuracy to {\it Gaia}, facilitating a dramatic improvement in turbulence-reduction modeling for \Rubn. Importantly, this strategy requires only catalog-level astrometric data from \Romn\ to re-reduce the pixel-level data from \Rubn.




See Table of Implementation Requirements:  \ref{cross-match}, \ref{sed},   \ref{astrometry}.





\subsection{Galaxy evolution and large-scale structure through cosmic time.}
\begin{figure}
    \centering
    \includegraphics[scale=0.4]{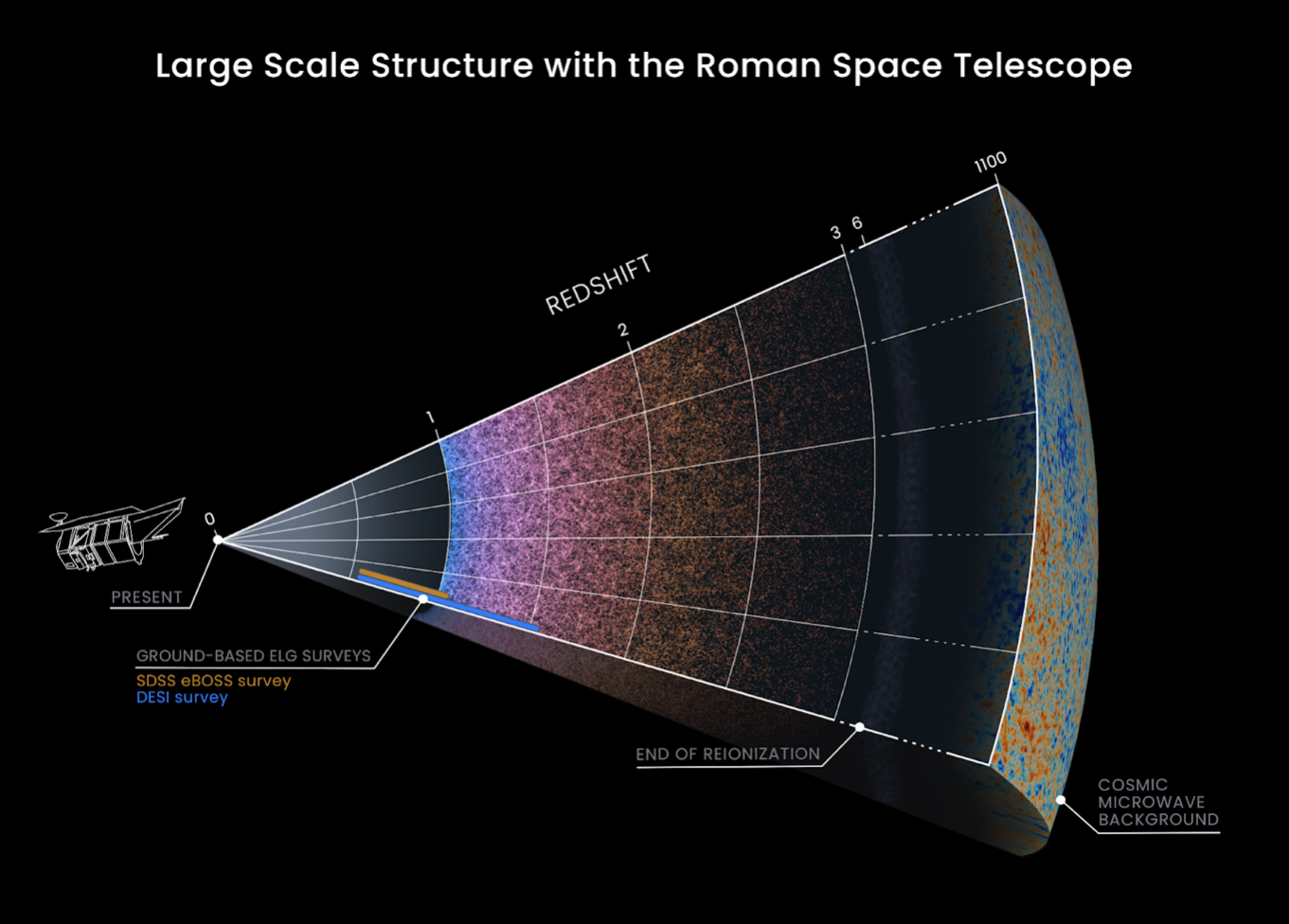}
    \caption{The \Romn\ High Latitude Wide Area Survey will measure spectroscopic redshifts of tens of millions of galaxies, offering a unique insight into the 3D distribution of large scale structure.  [Image credit:  Data provided by Z. Zhai and Y. Wang (Caltech/IPAC); Data visualization by J. DePasquale and D. Player (STScI).  Image from \url{https://roman.gsfc.nasa.gov/high_latitude_wide_area_survey.html} .  \Rubn\ survey strategy is still under consideration, but is expected to include a complementary and overlapping survey (See https://pstn-051.lsst.io/ for the March 2021 \Rubn\ Survey Cadence report.)  }
    \label{fig:high_lat}
\end{figure}

There is tremendous cosmological information available in the 3D distribution of galaxies, and together, \Romn\ and \Rubn\ will produce unprecedented, detailed maps of the sky.  Photometric and spectroscopic galaxy surveys (including galaxy positions, redshifts, fluxes, and shapes) produced by these two observatories will offer new insight on cosmological models and the formation of large scale structure (Figure \ref{fig:high_lat}).


The astronomical community is interested in a breadth of science cases that probe large scale structure and galaxy evolution.  These include:

{\it Clusters \& protoclusters} -- \Rubn\, with its wider field and bluer imaging, probes a slightly different population of galaxy clusters than \Romn\, which has higher resolution and redder imaging. As a result, \Romn\ and \Rubn\ will probe complementary redshift ranges ($z=0.1-0.2$ and $z=0.5-0.7$, respectively).  
At higher redshifts, \Rubns\ survey plus \Romns\ spectroscopic follow-up can be useful to identify and characterize galaxy composition of proto-clusters [Ettori pitch].

{\it Reionization} -- The combined wavelength coverage of data from \Romn\ and \Rubn\, in particular \Romns\ near-infrared sensitivity and \Rubns\ variability monitoring capabilities, will allow us to select well-controlled samples of $z>6$ sources.
The nature and the global properties of these high-redshift sources, including galaxies and AGNs, such as the luminosity functions, star-formation rate densities, or the clustering properties, would be derived with unprecedented accuracy.  The large survey footprint of the combined survey also benefits from sampling a variety of environment conditions, and allowing us to explore how different galaxy populations contribute to cosmic reionization. [Pello pitch].  A combined survey could also extend our ability to identify quasars out to $z=7-10$, which would allow us to further map cosmic reionization.  [Tee pitch]

{\it Low surface brightness Universe} -- To study the low surface brightness Universe statistically, one needs a large area, deep survey.  \Romns\ high spatial resolution for object recognition and deblending, when combined with \Rubns\ unprecedented depth for finding faint sources, can provide the observations to explore this.  In this case, having high-quality photometric redshifts will be crucial because objects will be too faint for spectroscopic follow-up. [Montes pitch]

{\it Lensing, morphologies, and distances} -- For many \Romn$+$\Rubn\ synergistic applications, having a robust calibration of \Rubn\ photometric redshifts with \Romn\ spectroscopy will be imperative [Wallman pitch].  Cosmological parallax may provide a new probe of cosmological geometry, but such a measurement requires a new lens sample for parallax measurements.  \Rubns\ wide survey and photometric redshifts, plus \Romns\ high resolution to model the lenses are necessary for this work.  [Pierce pitch]




See Table of Implementation Requirements:    \ref{south},   \ref{cross-match}, \ref{errors},  \ref{specz},  \ref{colocate}, \ref{simulation},  \ref{train}.


\subsection{Data-driven discoveries with \Romn\ and \Rubn : Supporting simulations as parallel data sets}

The \Romn\ and \Rubn\ telescopes will generate orders of magnitude more data than previous missions. Rare astronomical objects will be discoverable in statistical quantities; common populations will achieve large enough sample sizes to be sensitive to second-order effects; previously undiscovered populations will be unveiled for investigation. The majority of these new discoveries are likely to be data-driven: based on characterizing these huge data sets using the advanced techniques of modern data science and machine learning rather than choosing a single model to explain the data. 

Examples of discoveries enabled by the data-driven approach include quantifying the number of dimensions (e.g., distinct physical parameters) needed to describe a data set, or identifying unusual objects that fall outside the manifolds that contain most of the data.  Such data-driven discovery requires cross-matched catalogs in addition to the ability to do cross-catalog queries (that is, searching for targets that meet some criteria that requires both \Romn\ and \Rubn\ data products, such as optical and near-infared time domain signatures or image products).  With large datasets like this, the only feasible way will be to co-locate the data sets.

The data-driven approach prescribes a fundamental role for simulations of the objects being studied. In particular, they are crucial to make quantitative comparisons with observations by multiple instruments, train machine learning models, deal with complex uncertainties, and connect observations in detail to the underlying physical processes that govern them. The intermediate step of synthetic observation---imprinting the error model of a particular instrument on the output of a simulation---is the critical yet underappreciated link between the excitement of discovery and the resolution of scientific questions. To fully achieve the scientific promise of \Romn\ and \Rubn\, investment in the development of simulations and synthetic observation pipelines for frontier science is important, as well as publicizing them among the community, training its members to use them, and determining what functionalities need to be added are all key to maximizing the potential of these synergies.

In many of the most exciting potential synergies between these two observatories, \Rubn\ acts as a discovery engine and \Romn\ as the follow-up instrument. In this mode, it is crucial to understand the impact of the selection function on the results of the initial object census and the subsequent selection of candidates for follow-up [Nadler pitch]. Simulated catalogs allow the quantification of these effects, extending the impact of the few yet valuable targeted observations relative to the full population and lending statistical power to generalizations. Furthermore, detailed simulations allow for forecasting of follow-up observations to explore sensitivity to unconstrained physics within the constraints of the initial census results, reconciling observations by very different instruments to a common underlying source. Realizing the full potential of synergistic observations with \Romn\ and \Rubn\ will require community access to a standardized, maintained suite of flexible instrument models, validated by the instrument teams, for carrying out synthetic observations.  

Enormous and synergistic data sets also pose new challenges for data exploration, which hinges on responsive tools for visualization. As one pitch put it, 

\noindent \emph{How do we most effectively leverage humans' remarkable abilities to detect patterns and the unexpected when one instrument may generate petabytes of data spanning imagery, spectra, diagnostic telemetry, catalogs, MCMC outputs, and machine learning data products?} (P.K.G. Williams, ``Visualizing Big and Wide Data in the 21st Century”)

\noindent While each mission will develop its own data exploration tools, realizing the full combined potential of \Romn\ and \Rubn\ will require a coordinated effort to build an environment where the data can be visualized and explored side-by-side. The more seamless the transitions between the different data products, the more power it will have to generate new discoveries.  

The vast size of the \Romn\ and \Rubn\ catalogs will allow many measurements to push to unprecedented sensitivity. This new regime comes with the challenge of coping with non-Gaussian uncertainties: for example in photometric redshifts, phase-space modeling, or stellar population modeling. Forward modeling with simulated data is the preferred tool for meeting these challenges, yet is currently extremely computationally demanding in almost all cases. New advances---in computational infrastructure, interpolation algorithms, and analytical statistics---are needed to speed up forward modeling of \Romn$+$\Rubn\ data for key science questions.






See Table of Implementation Requirements:   \ref{colocate},  \ref{simulation}, \ref{train}, \ref{forward}.  See also \citet{chary_jsp} for joint pixel-level analysis infrastructure recommendations.

\begin{acknowledgements}
{\it Acknowledgements}: We would like to thank the ex-officio members of our \Romn\ and \Rubn\ Synergy working group for their guidance and input throughout this process, including Harry Ferguson (STScI), \v{Z}eljko Ivezi\'{c} (University of Washington), Neill Reid (STScI), and Beth Willman (AURA).
\end{acknowledgements}

\section{Table of Implementation Requirements}

\begin{longtable}{|c|p{15cm}|}

\hline
 & {\it Observing Strategy} \\
\hline

 \namedlabel{non-sidereal}{1.1} & Non-sidereal tracking by \Romn\ and \Rubn\ \\

 \namedlabel{too_roman}{1.2} & Target of opportunity follow-up imaging with \Romn\ \\
 
 \namedlabel{too_roman_spec}{1.3} & Target of opportunity follow-up spectroscopy with \Romn\ \\

 \namedlabel{too_rubin}{1.4} & Target of opportunity follow-up imaging with \Rubn\ \\

 \namedlabel{ecliptic}{1.5} & Survey observations of the ecliptic by \Romn\ and \Rubn\\\

 \namedlabel{rges_rubin}{1.6} & $\Rubn$ observations of the \Romn\ Galactic Bulge Time Domain Survey to fill in gaps between the observing windows and extend the baseline of the observations, and provide contemporaneous (within 1 day) observations\\

 \namedlabel{rges_rubin_pre}{1.7} & $\Rubn$ precursor observations of the \Romn\ Galactic Bulge Time Domain Survey fields to characterize potential microlensing source stars, and provide an astrometric and photometric baseline in order to reliably detect microlensing events\\

\namedlabel{bulge_rubin}{1.8} & Survey observations of the Galactic Plane and Bulge by \Rubn\ in its Deep Wide Fast Survey\\

\namedlabel{plane_roman}{1.9} & $\Romn$ Early Definition Astrophysics survey of 1000 deg$^2$ of the Galactic bulge/plane footprint \\

 \namedlabel{GI_roman}{1.10} & Guest Investigator programs for follow-up imaging by \Romn\ \\

 \namedlabel{GI_roman_spec}{1.11} & Guest Investigator programs for follow-up spectroscopy by \Romn\ \\

 \namedlabel{GI_rubin}{1.12} & Guest Investigator programs for follow-up observations by \Rubn\ \\

 \namedlabel{south}{1.13} & Choose Southern Hemisphere fields for the the \Romn\ High Latitude Wide Field Survey and High Latitude Time Domain Survey to overlap with \Rubn, including the \Rubn\ Deep Drilling Fields\\

 \namedlabel{early_astrometry}{1.14} & Complete one sweep of the \Romn\ High Latitude Wide Field Survey within the first year of operations to maximize astrometry potential\\

\namedlabel{roman_wide}{1.15} & Augment the \Romn\ High Latitude Survey to include a slow (1 month cadence) time domain survey\\

\namedlabel{rubin_roman_tda}{1.16} & $\Rubn$ Deep Wide Fast Survey observations of the \Romn\ Time Domain Survey fields\\

 \namedlabel{roman_cluster}{1.17} & $\Romn$ Early Definition Astrophysics survey of nearby galaxy groups and clusters \\
 
 \hline
 & {\it Data  Products} \\
 \hline
 
 \namedlabel{roman_mo}{2.1} & Automated identification and specialized image processing for moving objects in the \Romn\ images \\

  \namedlabel{mpc}{2.2} & Timely reporting of moving object detections by \Romn\ and \Rubn\ to the Minor Planet Center \\

  \namedlabel{cross-match}{2.3} & Cross-matched catalogs, joint source detection, cross-survey forced photometry, and joint astrometric solutions \\

  \namedlabel{cross-match-moving}{2.4} & Cross-matched catalogs, joint source detection, cross-survey forced photometry for moving objects \\

  \namedlabel{alerts_roman}{2.5} & Timely reporting of transient alerts by \Romn\ to the community \\

  \namedlabel{ml_roman}{2.6} & Machine-learning classification of morphology on \Romn\ data to identify structures such as strongly lenses systems \\

  \namedlabel{watchlist_rubin}{2.7} & The ability to create a watchlist of interesting targets to be notified when \Rubn\ detects a spatially coincident transient alert \\

  \namedlabel{errors}{2.8} & Detailed understanding of instrumental effects, selection functions, and sources of systematic errors for both \Romn\ and \Rubn\ \\

\hline
 & {\it Joint Analysis Methods} \\
\hline 

  \namedlabel{precovery}{3.1} & Precovery of newly identified \Rubn\ targets in earlier serendipitous \Romn\ observations \\

  \namedlabel{sed}{3.2} & Joint value-added catalogs of static targets (deblending, photo-z’s, morphology, colors, SED fitting) \\
  
  \namedlabel{forced}{3.3} & Forced photometry, including ability to forced photometer subsets of the data like individual epochs e.g., for moving (solar system) objects \\

  \namedlabel{forced_spec}{3.4} & Forced spectroscopic extraction in \Romn\ using \Rubn\ imaging as input catalog including moving objects \\

  \namedlabel{astrometry}{3.5}  & Astrometric cross-calibration of \Rubn\ pixel-level data with turbulence-reduction modeling using \Romn\ catalogs to enable precision proper motion measurements \\

  \namedlabel{cross_galaxy}{3.6} & Catalogs of key galaxy parameters constructed from joint $\Romn$+\Rubn\ catalogs should be made early, such that alert brokers can incorporate this information when processing new transients \\

  \namedlabel{specz}{3.7} &  $\Romn$ spectroscopic calibration of \Rubn\ photo-z’s \\

\hline
 & {\it Computing Infrastructure} \\
 \hline

  \namedlabel{comm_mma}{4.1} & Coordination and communication of \Romn\ and \Rubn\ with multi-messenger facilities, with a clear set of trigger criteria for TOO observations, and a clear (automated) implementation plan \\

  \namedlabel{colocate}{4.2} & Co-location of \Romn\ and \Rubn\ data in a common analysis environment to bring compute to the data \\

  \namedlabel{simulation}{4.3} & Support for simulated data: execution and documentation of a simulation database, up to date (versioned) instrument models for synthetic observations, parallel analysis of real and simulated data, catalog-level error modeling \\

  \namedlabel{train}{4.4} & Train the community in recent computational advances in big data (beyond querying databases, e.g., cloud computing) \\

  \namedlabel{forward}{4.5} & New advances---in computational infrastructure, interpolation algorithms, and analytical statistics---are needed to speed up forward modeling of $\Romn$+\Rubn\ data \\
\hline






  
  \hline 
\end{longtable}

\newpage 

\section{Appendix: Community Science Pitches}

\begin{longtable}{|p{5cm}|p{10cm}|}
\hline
{\bf First Author} & {\bf Title of Community Science Pitch} \\
\hline

\multicolumn{2}{|c|}{\bf MILKY WAY AND SOLAR SYSTEM SCIENCE} \\
 \hline 
Abrams, Natasha & {\it Using microlensing distributions to perform Galactic population analysis.}  \\
\hline

Busa, Innocenza  & {\it Uncovering new sources of Galactic cosmic rays from coordinated optical and near-infrared monitoring of diffuse clouds in the Galactic Bulge.}  \\
\hline

Chakrabarti, Sukanya  & {\it Direct accelerations measurements of eclipsing binaries to probe dark matter in the Milky Way.}  \\
\hline
Couperus, Andrew & {\it $\Rubn$ observations of the cycles of host stars of $\Romn$ detected exoplanets.}  \\
\hline

De, Kishalay & {\it Accreting compact objects in the dynamic Galactic plane.
What are the demographics of accreting compact objects in the Milky Way?} \\
\hline 

Eggl, Siegfried  & {\it Joint analysis of Solar System objects from $\Romn$ and $\Rubn$ observations of the ecliptic} \\
\hline

Fern\'{a}ndez-Trincado,~Jos\'{e} G. & {\it Dusting-off the Ancient Relics from the heart of the Milky Way.  What is the nature of the Milky Way's globular clusters?}  \\
\hline

Ginsburg, Adam &  {\it $\Romn$ and $\Rubn$ observations of young stellar object (YSO) outbursts, and the identification of obscured YSO populations.} \\
\hline
Jedicke, Rob & {\it Characterizing Extreme Members in $\Rubns$ Solar System Catalogue.  What is the composition, physical state, and morphology of the most unusual small bodies discovered by the LSST?} \\

\hline

Kastner, Joel  & {\it How do pre-main sequence stars obtain their final masses?   Rapid response Roman observations of $\Rubn$-detected eruptions of Young Stellar Objects.} \\
\hline

Lam, Casey & {\it Uncovering the hidden isolated black hole population: How many isolated black holes are there in the Milky Way?} \\
\hline

Lu, Jessica & {\it Finding primordial black holes with microlensing.}  \\
\hline
Meisner, Aaron & {\it A new window into the Milky Way’s structure and history.  What is the nature of star formation at very low masses?} \\
\hline

Minniti, Dante & {\it Old Populations in the Innermost Milky Way: A $\Romn$ and $\Rubn$ search for RR Lyrae in the Galactic Center.}  \\
\hline

Sainio, Arttu &  {\it $\Romn$ observations of the coldest brown dwarfs and rogue exoplanets.}	\\
\hline

Soares-Furtado, Melinda & {\it  Searching for moons around rogue planets.  What percentage of these planetary-mass objects harbor transiting companions, and what are the architectures of these bounded systems?} \\
\hline

Street, Rachel & {\it Characterizing and enhancing the known population of cold exoplanets.  What are the masses and dynamics of free-floating planets?} \\
\hline

Wyrzykowski, Lukasz  & {\it Astrometric microlensing signal from black holes.} \\
\hline 

Zucker, Catherine & {\it Mapping the Milky Way with a $\Romn$+$\Rubn$ Galactic Plane survey.} \\

\hline
\multicolumn{2}{|c|}{\bf TRANSIENT DISCOVERY AND TRANSIENT HOSTS} \\
\hline

Bentz, Misty &
{\it The morphologies and environments of galaxies hosting changing-look AGN. What is the physical mechanism behind rapid and large changes in the accretion rates of active galactic nuclei?} \\
\hline
Berger, Edo &  {\it Optical-infrared follow-up of gravitational wave events.  What are the electromagnetic properties of gravitational wave events, their relation to binary properties, and implications for r-process nucleosynthesis?} \\
\hline

Chase, Eve & {\it $\Romn$ and $\Rubn$ detection of kilonovae from follow-up of short-duration gamma-ray bursts or a gravitational-wave detections.}  \\
\hline

Coughlin, Michael & {\it Identifying and Characterizing the Sources of $r$-process material in $\Rubn$'s alert stream.  What fraction of heavy elements do neutron star mergers produce?} \\

\hline
Fox, Ori & {\it Probing the first stars, the evolution of the early Initial Mass Function (IMF), and the sources of reionization with transient astronomy in the early, high-redshift ($z > 6$) universe.} \\

\hline

French, Decker & {\it a) Measuring the Black Hole - Bulge relation across cosmic time.  How do galaxies and supermassive black holes co-evolve?  b) Connecting transients to their host galaxies.  How can we best identify large numbers of transient events?} \\
\hline

Gezari, Suvi & {\it Pinpointing the environments and hosts of $\Rubn$-discovered transients with $\Romn$ imaging.  What are the progenitors and/or central engines of cosmic explosions?} \\

\hline
Holwerda, Benne & {\it Using lensed supernova observations as an independent measure of $H_0$.} \\
\hline
Holz, Daniel & {\it EM follow-up of GW sources.} \\
\hline

Homayouni, Yasaman & {\it Census of black hole mass and accretion disks with $\Romn$-$\Rubn$.}  \\

\hline

Kilpatrick, Charlie & {\it Early $\Romn$ survey of nearby galaxy groups and clusters in the $\Rubn$ LSST footprint, especially Fornax and Virgo.  Systematically addressing massive star evolution and death.} \\
\hline

Lacy, Mark & {\it Quasar/AGN host galaxy studies in the era of the Nancy Grace $\Romn$ Space Telescope and Vera C. $\Rubn$ Observatory.}  \\
\hline

Liu, Xin & {\it Sub-arcsec quasar pairs and lenses from combining $\Rubn$ and $\Romn$.}  \\
\hline

Lyu, Jianwei & {\it Infrared reverberation mapping of active galactic nuclei.}  \\
\hline

Matsuura, Mikako & {\it How common is dust formation in supernovae?} \\
\hline
Mattila, Seppo & {\it Systematic study of dust emission in $\Rubn$-discovered transients with $\Romn$ imaging.}  \\

\hline

Perley, Dan & {\it Unusual and obscured transients at high redshift. What populations of transients remain unseen?} \\
\hline

Shen, Yue & {\it a. The discovery of a large statistical sample of $\sim$kpc-scale dual quasars from $\Romn$+$\Rubn$ to improve our understanding of the pairing and evolution of binary SMBHs in galaxy mergers.  b. Measuring the time lags from the dusty torus in AGN.} \\
\hline
Villar, V.~Ashley & {\it $\Romn$+$\Rubn$ observations of a $\Rubn$ Deep Drilling Field to finally unveil the late-time evolution of stripped-envelope core-collapse supernovae.}\\

\hline
Vinko, Jozsef & {\it Separating low- and high-redshift supernovae in a cosmological survey with $\Romn$.  How does the cosmic star formation rate behave at very high ($z > 6$) redshifts?}  \\
\hline

Windhorst, Rogier & {\it Observing the first stars directly with $\Rubn$ and $\Romn$ through cluster caustic transits.}  \\
\hline

Zhang, Jielai & {\it Kilonovae: A new window into fundamental physics, element creation, transient astronomy, and cosmology.  What is the equation of state of dense dense matter, and are kilonovae responsible for producing more than half the elements heavier than iron?}  \\
\hline

\multicolumn{2}{|c|}{\bf GALAXIES, LARGE SCALE STRUCTURE, AND DARK MATTER}  \\
\hline

Bechtol, Keitha & {\it $\Romn$ and $\Rubn$ coordination to enhance astrophysical probes of dark matter.} \\
\hline

Bell, Eric & {\it Reshaping our understanding of small-scale cosmology and the hierarchical growth of galaxies in the Local Universe.}  \\
\hline
Belokurov, Vasily & {\it Detecting and characterising Galactic halo sub-structure with $\Rubn$ and $\Romn$} \\
\hline

Birrer, Simon & {\it Strong lensing tomography with joint $\Rubn$+$\Romn$ deblending of strongly lensed arcs.  What is the geometry of the Universe?}  \\
\hline
Blain, Andrew & {\it	The role of large-scale structure in controlling the morphology of the zoo of galaxies.} \\
\hline

Dell'Antonio, Ian & {\it Studying the structure and populations of galaxy clusters with $\Romn$ and $\Rubn$.}  \\
\hline

Ettori, Stefano & {\it Proto-clusters and first collapsed structure at z>2.  Identifying and characterizing the galaxy composition of the first structures that form on Mpc scales (with masses >1 $\times 10^{13} M_\odot$) using $\Rubn$ imaging and $\Romn$ dedicated spectroscopic follow-up.} \\

\hline

Laine, Seppo & {\it Power of multiwavelength SEDs in deciphering stellar stream progenitors.  What are the stellar population metallicities, masses, 
and star formation histories of stellar stream progenitors?}  \\
\hline

Masters, Daniel & {\it A complete picture of galaxy evolution driven by the unparalleled statistical power of $\Romn$ and $\Rubn$ photometric data.}  \\
\hline

Montes, Mireia & {\it Exploring the low-surface brightness universe statistically with $\Romn$ and $\Rubn$.} \\
\hline

Nadler, Ethan	& {\it How does dark matter cluster on the smallest observationally accessible scales, and what does this teach us about dark matter particle properties?}  \\
\hline

Ntampaka, Michelle & {\it Enabling modern data analysis methods.  $\Romn$ and $\Rubn$ will produce exquisitely detailed observations - how can our data analysis tools rise to the challenge and use these data to their fullest potential?} \\
\hline

Patel, Ekta & {\it Probing the faint end of the galaxy luminosity function beyond the Local Group.
How do complete censuses of ultra-faint dwarf galaxies (to $M_V > -7$) beyond the Local Group improve our understanding of small-scale structure?} \\
\hline

Pello, Roser & {\it Towards a complete census of ionizing  sources during the first billion years.  What is the relative contribution of the different galaxy populations to cosmic reionization?} \\
\hline 

Pierce, Michael & {\it Measuring cosmological parallax with gravitationally lensed quasars and galaxies.}  \\
\hline

Rawls, Meredith & {\it Quantifying the impact of low-Earth-orbit satellite proliferation on ground-based astronomy.  How do increasing numbers of bright commercial satellites impact $\Rubn$'s science programs?} \\
\hline
Scognamiglio, Diana & {\it Relics with $\Rubn$ \& $\Romn$ ($R^3$): constraining the assembly and cosmic evolution of the most massive galaxies in the Universe.  How do the most massive galaxies form? How do they evolve in cosmic time?
} \\
\hline
Shivaei, Irene & {\it The effect of environment on galaxy evolution processes at Cosmic Noon ($z \sim 1-3$).
How does galaxies morphologies, their distribution of star formation, stellar mass, and dust varies across environment?
What is the connection between the AGN and their host galaxies' morphologies, stellar populations, and environment?} \\
\hline 
Simon, Josh & {\it Stellar archaeology in ultra-faint dwarf galaxies.}  \\
\hline
Tee, Wei Leong, & {\it The first quasars in the Universe: $\Romn$+$\Rubn$ synergy.} \\

\hline

Tran, Kim-Vy & {\it Identifying strong lensing candidates with machine-learning on $\Rubn$ and modeling the lenses and sources with $\Romn$ high resolution imaging and grism spectroscopy.}   \\ 
\hline

Wallman, Bruce & {\it Calibration of $\Rubn$ photometric redshifts with $\Romn$ spectroscopy} \\
\hline
Williams, Benjamin & {\it Fundamental improvement of our observational constraints on the structure and stellar content of the halos of nearby galaxies.}  \\
\hline

Williams, Peter K.~G. & {\it Visualizing big and wide data in the 21st Century.  What are the visualization tools that will empower astronomers to get the most science out of 21st-century datasets?} \\

\hline

Yoon, Ilsang &	{\it Evolution of the rest-frame optical galaxy structure and environmental effect up to z=2.5.} \\
\hline
\end{longtable}


\bibliography{r2d2.bib}{}
\bibliographystyle{aasjournal}

\end{document}